\documentclass{elsartL}
\usepackage{graphicx}
\usepackage{amssymb}
\usepackage{amsmath}
\usepackage{mathrsfs}
\usepackage{mathtools}
\usepackage{physics}
\usepackage{wasysym} 
\usepackage{mathtools}
\newcommand{\avg}[1]{\left< #1 \right>} 
\DeclarePairedDelimiter\floor{\lfloor}{\rfloor}
\begin{document}

\begin{frontmatter}
\title{Analysis of luminosity measurements of the pre-white dwarf PG 1159-035}
\author{E.~Matsinos{$^*$}}

\begin{abstract}The study of the luminosity measurements of the pre-white dwarf PG 1159-035 has established the properties of the rich power spectrum of the detected radiation and, derived thereof, the physical properties of 
this celestial body. Those of the measurements which are available online are analysed in this work from a different perspective. After the measurements were band-passed, they were split into two parts (of comparable sizes), 
one yielding the training (learning) set (i.e., the database of embedding vectors and associated predictions), the other the test set. The optimal embedding dimension $m_0=10$ was obtained using Cao's method; this result was 
confirmed by an analysis of the correlation dimension. Subsequently, the extraction of the maximal Lyapunov exponent $\lambda$ was pursued for embedding dimensions $m$ between $3$ and $12$; results were obtained after removing 
the prominent undulations of the out-of-sample prediction-error arrays $S (k)$ by fitting a monotonic function to the data. The grand mean of the values, obtained for sufficient embedding dimensions ($10 \leq m \leq 12$), was: 
$\lambda = (9.2 \pm 1.0 ({\rm stat.}) \pm 2.7 ({\rm syst.})) \cdot 10^{-2}~\Delta \tau^{-1}$, where $\Delta \tau=10$ s is the sampling interval in the measurements. On the basis of this significantly non-zero result, it may be 
concluded that the physical processes, underlying the variation of the luminosity of PG 1159-035, are non-linear. The aforementioned result for $\lambda$ was obtained using the $L^\infty$-norm distance; a larger, yet not 
incompatible, result was extracted with the Euclidean ($L^2$-norm) distance.\\
\noindent {\it PACS 2010:} 05.10.-a; 05.45.-a; 05.45.Gg; 45.30.+s; 95.10.Fh
\end{abstract}
{$^*$}{Electronic mail: evangelos (dot) matsinos (at) sunrise (dot) ch}
\end{frontmatter}

\section{\label{sec:Introduction}Introduction}

When the hydrogen reserves of a star are exhausted, the star collapses until the temperature in its core enables helium to ignite and burn to carbon. To determine the fate of the star, one only needs to know its mass. For 
stellar masses comparable to the Solar mass ($M_{\astrosun}$), the outer shells of the star evaporate into space (planetary nebula) leaving a relic at the centre, which evolves into a white dwarf. Discovered in 1977 in a 
project aiming at the identification of ultraviolet-excess stellar sources (Palomar-Green survey) \cite{green1986}, PG 1159-035 is a celestial body in the Constellation of Virgo, which the experts in the domain of Stellar 
Evolution place in a transitional phase, from the central star of a planetary nebula to a white dwarf. PG 1159-035 contains about $60 \%$ of the solar mass, confined within a radius of about $2.5 \%$ of the solar radius 
(about $2.7$ times the radius of the Earth), and is a few hundred times more luminous than our Sun. Stellar bodies in this transitional phase are known as `pre-white dwarves'.

The variation of the luminosity of the pre-white dwarf PG 1159-035, arising from non-radial gravity-wave ($g$-wave) pulsations, was measured (for the first time to that detail) by the Whole Earth Telescope (WET) in 1989. The 
members of the Collaboration introduce the WET project as ``a global, interactive network of photometric observers who together provide essentially continuous coverage of a set of prioritised targets.'' \cite{winget1991} The 
pre-white dwarf PG 1159-035 might have attracted original attention because its ``near-equatorial declination allowed observatories in both hemispheres to participate in the observations.''

The analysis of the measurements, spanning over $229$ effective hours (hr) of data acquisition, revealed $122$ pulsation modes, with periods ranging from $300$ to $1000$ s \cite{winget1991}. Those of us who do not frequently 
come across Fourier transforms with a resolution of $1~\mu$Hz will undoubtedly be impressed by Fig.~4 of Ref.~\cite{winget1991}. That work set the region of interest in the power spectrum of the detected light (p.~330): ``The 
peaks of greatest power in the Fourier spectrum are largely confined to the interval between $1000$ and $2600~\mu$Hz, with the dominant power in the narrower interval between $1750$ and $2250~\mu$Hz.'' Reference \cite{winget1991} 
was also important for another, less obvious reason: it cemented the foundations of the domain of Asteroseismology, as the discipline studying the ``stellar structure and evolution as revealed by global stellar oscillations.'' 
To provide an idea of what may be learnt from the analysis of such observations, the authors of Ref.~\cite{winget1991} write (about PG 1159-035) in the abstract of their paper: ``We find its mass to be $0.586~M_{\astrosun}$, 
its rotation period $1.38$ days, its magnetic field less than $6000$ G, its pulsation and rotation axes to be aligned, and its outer layers to be compositionally stratified.''

The entirety of the luminosity measurements of PG 1159-035 (runs between 1979 and 2002) were analysed in a more recent paper \cite{costa2008}, leading to the detection of $76$ additional (i.e., on top of those which had been 
identified in Ref.~\cite{winget1991}) pulsation modes. As the authors mention in the abstract of their paper, the $122+76=198$ known pulsation modes of PG 1159-035 represent ``the largest number of modes detected in any star 
besides the Sun.'' Furthermore, Ref.~\cite{costa2008} improved on the accuracy of the physical quantities of PG 1159-035, e.g., on its rotation period ($1.3920 \pm 0.0008$ days), and provided updated estimates for the object's 
mass ($0.59 \pm 0.02~M_{\astrosun}$) and magnetic field ($<2000$ G).

A number of models, aiming at the description of the measurements obtained from PG 1159-035 and from similar stars in terms of the physical processes in the interior of these bodies, may be found in the literature 
\cite{kawaler1994,garciaberro2006,corsico2006,corsico2008,althaus2008,costakepler2008}. The free parameters of these models include the stellar mass, the effective temperature $T_{\rm eff}$, the surface helium-layer thickness 
(usually expressed as a fraction of the stellar mass), and the stellar composition. Using such models, knowledge may be gained of the temporal evolution of celestial bodies which have split off from the so-called Post-Asymptotic 
Giant Branch, a horizontal branch in the Hertzsprung-Russell diagram, characterised by nearly constant luminosity and sharply decreasing temperature. The temporal evolution of these stars follows mass-dependent curves in the 
($\lg T_{\rm eff}$,$\lg g$) plane\footnote{Of course, $\lg x \coloneqq \log_{10} x$.}, where $g$ stands for the ratio of the gravitational acceleration at the surface of the star to that at the surface of the Earth.

By variation of the free parameters of the aforementioned models, stellar solutions are obtained and allowed to evolve in time. A database of stellar states (snapshots in the evolution of each simulated body) is thus created. 
After the (time-dependent) predictions for the pulsation spectra are obtained in each of these solutions, the database element, which best resembles the pulsation spectra obtained from the luminosity measurements of a given 
source, may be identified. Reliable information on the progenitor of the specific object, as well as forecasts for its future, may thus be obtained.

References \cite{kawaler1994,garciaberro2006,corsico2006,corsico2008,althaus2008,costakepler2008} look at each such celestial body from the point of view of Physics, i.e., attempting its description in terms of the established 
principles of Astrophysics and, in particular, of Stellar Evolution. This work looks at the acquired data from a different perspective, investigating the possibility that the data alone could provide an answer on whether the 
physical processes, generating the observations, are linear or non-linear. Knowledge of the physical system is used as input only in the filtering of the scalar time-series measurements, namely in setting the appropriate 
band-pass/stop characteristics, as they have been known from Refs.~\cite{winget1991,costa2008}. To the best of my knowledge, this is the first study of the time-series measurements of PG 1159-035 from the perspective of 
non-linear dynamics.

This work uses those of the time-series measurements of PG 1159-035, which may be found (among other data from a variety of scientific domains) in the web site \cite{timeseries} (data set E); henceforth, these measurements 
will be referred to as `original' (though, in reality, they have been selected from a larger set of data). Reference \cite{timeseries} does not provide information on how the available data has been obtained from the set of 
measurements acquired in the WET 1989 runs.

The software development, relating to this work, is part of a broader and more ambitious programme, aiming at robust analyses of time series via the application of user-selected linear and/or non-linear methods. Free software 
performing such analyses has been available since a long time, e.g., see Ref.~\cite{hegger2007}.

\section{\label{sec:Data}The original time-series arrays}

Seventeen data sets, containing luminosity measurements of PG 1159-035 from the 1989 runs, may be found in the web page \cite{timeseries}. These data sets carry the prefix `SF\_E\_', which will be replaced here by `E'. The 
total number of measurements in these files is $27191$, corresponding to an effective time of $271910$ s. On the other hand, more measurements are shown\footnote{In the caption of Fig.~1 of Ref.~\cite{winget1991}, it is 
mentioned that the data shown therein corresponds to the central $6$ days of the run. Therefore, one might be led to conclude that the displayed data represent a fraction of the luminosity measurements collected in the WET 
project in the 1989 runs.} in Fig.~1 of Ref.~\cite{winget1991}. Furthermore, from Table 1 of Ref.~\cite{costa2008} one extracts the information that a total of $82471$ measurements had been acquired in the 1989 runs, i.e., 
about three times the amount of measurements found in Ref.~\cite{timeseries}. As a result, it is not easy to determine which parts of the data, acquired in the 1989 runs, found their way into the snippets of measurements 
contained in the database of Ref.~\cite{timeseries}. Some statistical information on the seventeen time series of Ref.~\cite{timeseries} is given in Table \ref{tab:RawData}.

\begin{table}
{\bf \caption{\label{tab:RawData}}}Quantities relating to the available luminosity measurements of PG 1159-035, obtained from Ref.~\cite{timeseries} (data set E therein). The columns correspond to an identifier of each 
time-series array, the number of its measurements ($N$), its minimal and maximal values ($s_{\rm min}$ and $s_{\rm max}$, respectively), the range of variation of its entries, the average, and the root-mean-square (rms) of 
the signal in the data set. In the data acquisition, the sampling frequency was $100$ mHz (hence the Nyquist frequency $f_N$ was $50$ mHz) and, naturally, the sampling interval $\Delta \tau$ was $10$ s. For the sake of example, 
the data set E01 contains continuous luminosity measurements spanning about $1$ hr and $42$ minutes.
\vspace{0.2cm}
\begin{center}
\begin{tabular}{|c|c|c|c|c|c|c|}
\hline
Data set & $N$ & $s_{\rm min}$ & $s_{\rm max}$ & $s_{\rm max}-s_{\rm min}$ & $\avg{s}$ & rms\\
\hline
\hline
E01 & $617$ & $-0.3153$ & $0.2904$ & $0.6057$ & $3.03 \cdot 10^{-4}$ & $1.27 \cdot 10^{-1}$\\
E02 & $1255$ & $-0.3053$ & $0.3398$ & $0.6451$ & $-4.65 \cdot 10^{-5}$ & $1.03 \cdot 10^{-1}$\\
E03 & $1221$ & $-0.2956$ & $0.2713$ & $0.5669$ & $1.02 \cdot 10^{-4}$ & $9.49 \cdot 10^{-2}$\\
E04 & $979$ & $-0.3290$ & $0.3149$ & $0.6439$ & $1.21 \cdot 10^{-4}$ & $1.08 \cdot 10^{-1}$\\
E05 & $549$ & $-0.3223$ & $0.2991$ & $0.6214$ & $-1.59 \cdot 10^{-4}$ & $1.03 \cdot 10^{-1}$\\
E06 & $1553$ & $-0.3163$ & $0.3401$ & $0.6564$ & $1.12 \cdot 10^{-4}$ & $1.08 \cdot 10^{-1}$\\
E07 & $1936$ & $-0.5947$ & $0.6315$ & $1.2262$ & $4.55 \cdot 10^{-5}$ & $1.73 \cdot 10^{-1}$\\
E08 & $2495$ & $-0.4549$ & $0.3760$ & $0.8309$ & $-7.09 \cdot 10^{-5}$ & $1.45 \cdot 10^{-1}$\\
E09 & $1940$ & $-0.6010$ & $0.5801$ & $1.1811$ & $5.78 \cdot 10^{-5}$ & $1.57 \cdot 10^{-1}$\\
E10 & $1471$ & $-0.4623$ & $0.4568$ & $0.9191$ & $6.43 \cdot 10^{-5}$ & $1.48 \cdot 10^{-1}$\\
E11 & $2605$ & $-0.5454$ & $0.6033$ & $1.1487$ & $1.19 \cdot 10^{-2}$ & $1.82 \cdot 10^{-1}$\\
E12 & $1548$ & $-0.5415$ & $0.3590$ & $0.9005$ & $6.36 \cdot 10^{-5}$ & $1.28 \cdot 10^{-1}$\\
E13 & $2568$ & $-0.3857$ & $0.3906$ & $0.7763$ & $-5.84 \cdot 10^{-7}$ & $1.17 \cdot 10^{-1}$\\
E14 & $2602$ & $-0.4517$ & $0.4022$ & $0.8539$ & $-1.65 \cdot 10^{-3}$ & $1.32 \cdot 10^{-1}$\\
E15 & $672$ & $-0.4322$ & $0.6539$ & $1.0861$ & $-3.38 \cdot 10^{-4}$ & $1.79 \cdot 10^{-1}$\\
E16 & $1512$ & $-0.5381$ & $0.6366$ & $1.1747$ & $5.22 \cdot 10^{-3}$ & $1.84 \cdot 10^{-1}$\\
E17 & $1668$ & $-0.2791$ & $0.3206$ & $0.5997$ & $1.33 \cdot 10^{-4}$ & $9.94 \cdot 10^{-2}$\\
\hline
\end{tabular}
\end{center}
\vspace{0.5cm}
\end{table}

\begin{figure}
\begin{center}
\includegraphics [width=15.5cm] {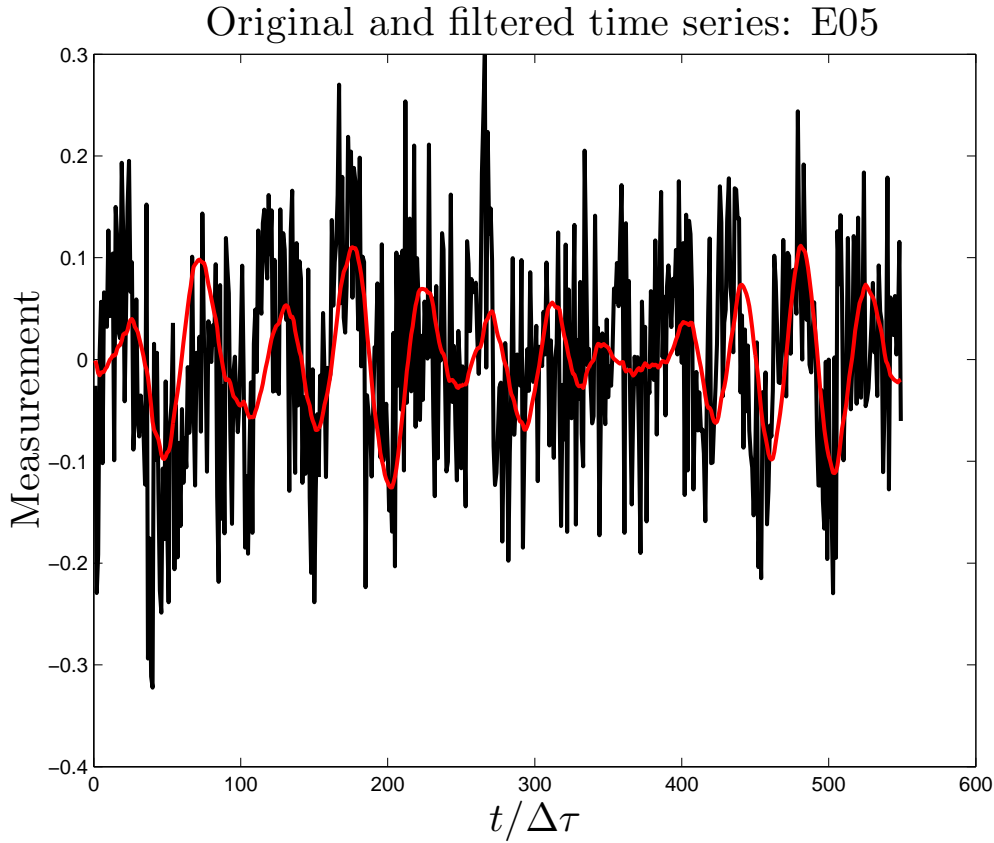}
\caption{\label{fig:Plot05}The luminosity measurements contained in the data set E05 of Ref.~\cite{timeseries}. The plot is shown as an extreme example of the amount of noise contained in the measurements of the pre-white dwarf 
PG 1159-035 available from Ref.~\cite{timeseries}. The measurements are connected with black line segments, whereas the filtered data (see Section \ref{sec:Filtering}) correspond to the red curve.}
\vspace{0.35cm}
\end{center}
\end{figure}

Difficulties with astronomical observations, in particular with those conducted via ground-based telescopy, are not infrequent; they may be caused by a variety of phenomena, predominantly atmospheric (turbulence, humidity, 
cloud coverage, etc.), but also relating to the position and the brightness of the Moon. Visual inspection of the luminosity measurements of PG 1159-035 reveals that, though an underlying oscillatory pattern in the intensity 
of the detected radiation is observed, a number of issues need to be resolved prior to the commencement of the data analysis. To start with, a large amount of noise seems to be present in the measurements. In this respect, the 
most extreme case appears to be the data set E05, shown in Fig.~\ref{fig:Plot05}. In addition to the obvious `high-frequency' noise, patterns in the data resemble those emanating from a drifting or oscillating calibration. There 
are cases in which a sizeable `drop of the values' appears around the middle of the data set, as in files E02, E15, and E16. Transient effects are also seen, e.g., over the first $100$ measurements in data set E01. Finally, the 
range of the signal appears to be close to the $0.6$ level for the first six data sets, considerably larger for the subsequent ten data sets, returning to the $0.6$ level for E17.

There are also places in the data sets where successive measurements appear to be in a perfect linear relationship, e.g., see files E02, E03, E04, E07, E09, E12, E14, and E17. One explanation of this effect may involve the 
removal of the background, as described on p.~327 of Ref.~\cite{winget1991}: ``\dots [W]e interrupted the observations of the target and comparison star for about $1$ minute of sky observations at regular intervals of roughly 
$20$ minutes (when the Moon was up and the sky was very bright) to $1$ hr (a more typical value since the bulk of the observations were obtained during dark-time). We then interpolated linearly between sky observations and 
subtracted the result from the data.''

\section{\label{sec:AnalysisOriginal}The analysis of the original time-series arrays}

The analysis of the time-series arrays of Ref.~\cite{timeseries} will generally follow the guidelines of the book `Nonlinear Time Series Analysis' by Kantz and Schreiber \cite{kantz1997}, a book which (in my opinion) should 
be of interest even to those who do not intend to perform non-linear analyses; for brevity, I will refer to these two authors as `KS' from now on. The stationarity of each time series of measurements is tested on the basis of 
the variation of the average and of the rms values of the measurements contained in successive data segments (of the given time series of measurements). The spectrogram is also analysed. If $N$ stands for the total number of 
measurements in a data set, $W$ for the length of the running windows (expressed in sampling intervals), $S \leq W$ for the shift of the window (also expressed in sampling intervals) in successive positions, $n$ for the number 
of windows, and $r$ for the total number of unpaired measurements (i.e., of the data points which are not included in any window), the relation arises:
\begin{equation*}
N = (n-1) S + W + r \, \, \, .
\end{equation*}
Therefore, $r = N - (n-1) S - W$. By varying $S$ and $W$ within user-defined limits ($5 \%$ limits are adopted in this work), starting from a $5 \%$ overlap ($d=1-S/W$) between successive windows, one may minimise $r$ separately 
for each time series, thus excluding (in the test of stationarity) as few measurements as possible. A few measurements are excluded only in the test of stationarity of the input time series, i.e., not in Section 
\ref{sec:AnalysisFiltered} where the dynamical invariants of the source of the observations are determined.

For the detection of the outliers in the sets of the average and of the rms values, Rosner's generalised ESD (Extreme Studentised Deviate) test \cite{rosner1983} was performed. In this algorithm, a set of $N_k$ values is tested 
for the presence of exactly $1$, $2$, \dots, $N_a$ outliers, where $N_a$ is a user-defined limit satisfying the condition $N_a < N_k$ (the test is performed for $N_a \leq \floor*{N_k/3}$). The advantage of Rosner's test (in 
comparison to other tests, e.g., to Grubbs's outlier test \cite{grubbs1969,stefansky1972}) is that the `optimal' number of outliers is obtained from the input data themselves. In addition, a $\chi^2$ test was performed on the 
set of the average values and their uncertainties (i.e., the standard error of the means) corresponding to each window: for a set of $n$ independent\footnote{Of course, unless the overlap between successive windows vanishes, 
the $x_i$ values are not independent. However, the average overlap between successive windows in this work was about $6.2 \%$, i.e., low enough to enable one to `gloss over' the issue of independence.} measurements $x_i$, with 
uncertainties $\delta x_i$, the $\chi^2$ value associated with the reproduction of the measurements by their weighted average is given by:
\begin{equation} \label{eq:EQ001}
\chi^2 = \sum_{i=1}^n w_i x_i^2 - \left( \sum_{i=1}^n w_i x_i \right)^2 / \sum_{i=1}^n w_i \, \, \, ,
\end{equation}
where the weight $w_i$ is equal to $(\delta x_i)^{-2}$. The comparison of the p-value, obtained from the $\chi^2$ of Eq.~(\ref{eq:EQ001}) and the number of degrees of freedom $n-1$ (one degree of freedom needs to be removed, 
as the constant of the reproduction of the input values is extracted from the values themselves), with a user-defined significance threshold $\mathrm{p}_{\rm min}$ enables the test of the constancy of the input values $x_i$: 
the constancy is accepted when $\mathrm{p} \geq \mathrm{p}_{\rm min}$, rejected when $\mathrm{p} < \mathrm{p}_{\rm min}$. In accordance with the choice of most statisticians, the $\mathrm{p}_{\rm min}$ threshold of $1.00 \cdot 10^{-2}$ 
was adopted herein as the outset of statistical significance.

I will next address some of the properties of the autocorrelation function\footnote{Although the term `autocorrelation function' is routinely used in the analysis of discrete time series, it is technically more appropriate to 
refer to the autocorrelation as a `sequence'. To retain simplicity however, I will also use the term `autocorrelation function' in this work, assume that a function is obtained from the sequence by interpolation, and refrain 
from introducing new terminology for the mere sake of formality.}. This function enables the extraction of an estimate for the embedding delay, which is one of the two important quantities in the reconstruction of the phase 
space (state space, for others) of the dynamical system yielding the time-series measurements, i.e., of the vector space within which the system `lives'. One important step when extracting estimates for the dynamical invariants 
of a non-linear system is the exclusion of the elements of the time series which are `temporally correlated' with a given element $s_k$; their closeness to $s_k$ is due to the frequency at which the dynamical system is being 
observed\footnote{To understand the removal of these temporally correlated elements, consider that, while driving your vehicle, you arrive at a four-way intersection (two perpendicular roads). Let me denote the four roads, 
starting at the intersection, as A, B, C, and D. The motion of the vehicle is defined if one sets the direction from which it reaches the intersection (e.g., C) and the one it follows after it (e.g., A). If one is not 
interested in providing estimates for the instantaneous velocity of the vehicle, all other observations (of the position of the vehicle as a function of time) are redundant.}. Such elements are not only superfluous in the 
context of time-delay embeddings; they actually obscure the geometrical features of the phase space. The determination of the embedding delay is the first step in a time-series analysis. In this work, the embedding delay will 
be expressed in sampling intervals: therefore, the embedding delay $\nu$ represents the temporal interval $\nu \Delta\tau$.

The second important quantity in time-delay embeddings is the dimension $m$ of the embedding vectors\footnote{In fact, it is pointed out in the literature that the most important parameter in time-delay embeddings is the 
\emph{time span of the embedding vectors}, namely the product $m \nu$ or equivalently the length of the embedding time window $m \nu \Delta\tau$. Nevertheless, I will follow the `traditional' approach herein, and regard $m$ 
and $n$ as the free parameters.}. Given an element with index $i > (m-1) \nu$ of the original time-series array $s_k$ ($k \in \{ 1, \dots, N \}$), the components of the embedding vectors ${\ss}_i$ are suitably chosen elements 
of the series, representing epochs which end at the time instant $i$:
\begin{equation} \label{eq:EQ002}
{\ss}_i \coloneqq (s_{i - (m-1) \nu}, s_{i - (m-2) \nu}, \dots, s_{i-\nu}, s_{i}) \, \, \, .
\end{equation}
Therefore, each time instant $i \Delta\tau$ may be associated with an $m$-dimensional vector with (for an appropriate choice for the lag) independent components. The study of these $m$-dimensional vectors ${\ss}_i$ enables the 
extraction of the important information, i.e., of the characteristics of the phase space of the dynamical system.

The unbiased variance of the measurements of a time series is given by the expression:
\begin{equation*}
\hat{\sigma}^2 = \frac{1}{N-1} \sum_{i=1}^N (s_i - \avg{s})^2 \, \, \, ,
\end{equation*}
where
\begin{equation*}
\avg{s} = \frac{1}{N} \sum_{i=1}^N s_i \, \, \, .
\end{equation*}
The unbiased variance at lag $j$ is defined as
\begin{equation*}
\hat{\sigma}_j^2 = \frac{1}{N-1-j} \sum_{i=j+1}^N (s_{i-j} - \avg{s}_j) (s_i - \avg{s}_j) \, \, \, ,
\end{equation*}
where the quantity $\avg{s}_j$ will be obtained from the variation of $\hat{\sigma}_j^2$. It may be proven that $\hat{\sigma}_j^2$ is minimal when
\begin{equation*}
\avg{s}_j = \frac{1}{2(N-j)} \sum_{i=j+1}^N (s_i + s_{i-j}) \, \, \, .
\end{equation*}
Finally,
\begin{equation*}
\hat{\sigma}_j^2 = \frac{1}{N-1-j} \left( \sum_{i=j+1}^N s_{i-j} s_i - (N-j) \avg{s}_j^2 \right) \, \, \, .
\end{equation*}
The autocorrelation at lag $j$ is defined as the ratio $\hat{\sigma}_j^2 / \hat{\sigma}_0^2 \equiv \hat{\sigma}_j^2 / \hat{\sigma}^2$. Therefore, the autocorrelation at lag $0$ is identically equal to $1$.

The autocorrelation function may provide information on the nature of the input time series. For instance, it is known that white noise has a random autocorrelation function, whereas signals from deterministic chaotic systems 
yield exponentially-decaying autocorrelation functions. KS favour the fixation of the embedding delay from the first root of the autocorrelation function (this option is also followed here). Another choice, presumably more 
suitable for the analysis of time series yielding slowly-decaying autocorrelation functions, would be to use the time at which the autocorrelation function drops (for the first time) below the value of $e^{-1}$. (Of course, 
there are other possibilities, e.g., to use the first minimum of the mutual-information function.) An evident question arises: if the emphasis is placed on the independence of the elements of the embedding vectors, why should 
one not simply choose an embedding delay considerably larger than the lag? The answer is that the choice of an `economical' embedding delay prevents the unfolding of the attractor (of a dissipative chaotic system) onto itself.

After these explanations, it is time I came to the results of the analysis of the seventeen original time-series arrays obtained from the pre-white dwarf PG 1159-035 (see Table \ref{tab:RawDataContinued}). The dominant frequency 
present in the measurements is equal to about $4 \%$ of the Nyquist frequency, i.e., about $2$ mHz, corresponding to a period of about $500$ s. To be able to test the stationarity of the input time series using an adequate 
amount of measurements\footnote{Regarding the tests of stationarity in a time series, Refs.~\cite{kantz1997} emphasise: ``\dots the time series should cover a stretch of time which is much longer than the longest characteristic 
time scale that is relevant for the evolution of the system.'' Although the definition of ``much longer'' is subjective, it is unlikely that the time-series arrays analysed in this work qualify as ``long enough'', and even 
more so the running windows used in the test of stationarity (which are only double the size of the dominant time scale). I would have preferred to use longer windows (e.g., $500$ sampling intervals), but this would not have 
worked on most of the snippets of Ref.~\cite{timeseries}.}, it was decided to make use of windows (initial value, subject to $5 \%$ variation in order to minimise $r$) spanning two periods of the dominant oscillation in the 
signal, covering about $1000$ s. This is the reason that the windows in Table \ref{tab:RawDataContinued} have a length of about $100$ sampling intervals.

Significant departure from stationarity is seen in all cases in Table \ref{tab:RawDataContinued}: there is no data set in which the stationarity of the signal can be asserted. In several cases, the Discrete Fourier Transform 
is enhanced at low frequencies, implying that the original time series exhibits long-term oscillations\footnote{On p.~629, the authors of Ref.~\cite{costa2008} put forward an explanation for the enhanced low-frequency (below 
$300~\mu$Hz), high-amplitude peaks in the Fourier spectrum, attributing their appearance to atmospheric phenomena, yielding corrections which are dependent on the colour of a star. As a result, a calibration obtained from a 
close-by (comparison) star of a different surface temperature (to the one of the target object) introduces spurious effects in the frequency spectrum of the object of interest.}; presumably, such effects do not have much to 
do with the system under observation. The long-term oscillations directly affect the autocorrelation function, leading to the extraction of longer embedding delays. (Of course, the enhanced low-frequency components in the power 
spectrum and the stationarity of the input data do not easily come to terms.) Inspection of Table \ref{tab:RawDataContinued} leaves little doubt that, in order to be of use, the available measurements must first be filtered.

\begin{table}
{\bf \caption{\label{tab:RawDataContinued}}}Some quantities relating to the analysis of the original luminosity measurements of PG 1159-035. $W$ is the length of the running window, $S$ its shift, and $\nu$ the embedding delay 
(first zero of the autocorrelation function); all these quantities are expressed in sampling intervals $\Delta\tau$. The quantity $d$ is the overlap between successive windows, $n$ is the number of windows, and $r$ stands for 
the number of unpaired points (i.e., for those which are not contained in any window); measurements are excluded only in the test of stationarity of each input time series. The column `p-value' contains the p-value pertaining 
to the test of stationarity (constancy of the average signal within each running window), obtained from the $\chi^2$ value of Eq.~(\ref{eq:EQ001}) and the number of degrees of freedom $n-1$. LFC stands for `Low-Frequency 
Components' and `DFT' for `Discrete Fourier Transform'.
\vspace{0.2cm}
\begin{center}
\begin{tabular}{|c|c|c|c|c|c|c|c|c|}
\hline
Data set & $W$ & $S$ & $d$ (\%) & $n$ & $r$ & p-value & $\nu$ & Comments\\
\hline
\hline
E01 & $105$ & $100$ & $5$ & $6$ & $12$ & $2.54 \cdot 10^{-3}$ & $14$ & Transient effects in first $100$ steps\\
\hline
E02 & $103$ & $96$ & $7$ & $13$ & $0$ & $2.34 \cdot 10^{-14}$ & $14$ & Trough in the middle of the data set\\
\hline
E03 & $105$ & $93$ & $11$ & $13$ & $0$ & $\approx 0$ & $15$ & \\
\hline
E04 & $105$ & $97$ & $8$ & $10$ & $1$ & $\approx 0$ & $18$ & \\
\hline
E05 & $99$ & $90$ & $9$ & $6$ & $0$ & $9.40 \cdot 10^{-3}$ & $13$ & No obvious structure in the data\\
\hline
E06 & $97$ & $91$ & $6$ & $17$ & $0$ & $8.66 \cdot 10^{-5}$ & $14$ & \\
\hline
E07 & $96$ & $92$ & $4$ & $21$ & $0$ & $\approx 0$ & $26$ & Enhanced LFC in DFT;\\
 & & & & & & & & slowly-decaying autocorrelation function\\
\hline
E08 & $103$ & $92$ & $11$ & $27$ & $0$ & $\approx 0$ & $15$ & \\
\hline
E09 & $97$ & $97$ & $0$ & $20$ & $0$ & $6.01 \cdot 10^{-14}$ & $14$ & \\
\hline
E10 & $99$ & $98$ & $1$ & $15$ & $0$ & $\approx 0$ & $15$ & \\
\hline
\end{tabular}
\end{center}
\vspace{0.5cm}
\end{table}

\newpage
\begin{table*}
{\bf Table \ref{tab:RawDataContinued} continued}
\vspace{0.2cm}
\begin{center}
\begin{tabular}{|c|c|c|c|c|c|c|c|c|}
\hline
Data set & $W$ & $S$ & $d$ (\%) & $n$ & $r$ & p-value & $\nu$ & Comments\\
\hline
\hline
E11 & $105$ & $100$ & $5$ & $26$ & $0$ & $\approx 0$ & $129$ & Enhanced LFC in DFT;\\
 & & & & & & & & oscillating autocorrelation function\\
\hline
E12 & $105$ & $90$ & $14$ & $17$ & $3$ & $\approx 0$ & $17$ & Enhanced LFC in DFT;\\
 & & & & & & & & peculiar peak in DFT at about $8.4$ mHz\\
\hline
E13 & $98$ & $95$ & $3$ & $27$ & $0$ & $\approx 0$ & $15$ & Enhanced LFC in DFT\\
\hline
E14 & $102$ & $100$ & $2$ & $26$ & $0$ & $6.43 \cdot 10^{-8}$ & $14$ & \\
\hline
E15 & $96$ & $96$ & $0$ & $7$ & $0$ & $\approx 0$ & $68$ & Enhanced LFC in DFT;\\
 & & & & & & & & oscillating autocorrelation function;\\
 & & & & & & & & trough in the middle of the data set;\\
 & & & & & & & & effects in last $100$ steps\\
\hline
E16 & $102$ & $94$ & $8$ & $16$ & $0$ & $\approx 0$ & $72$ & Enhanced LFC in DFT;\\
 & & & & & & & & oscillating autocorrelation function;\\
 & & & & & & & & trough in the middle of the data set\\
\hline
E17 & $104$ & $92$ & $12$ & $18$ & $0$ & $\approx 0$ & $19$ & Enhanced LFC in DFT\\
\hline
\end{tabular}
\end{center}
\vspace{0.5cm}
\end{table*}

\section{\label{sec:Filtering}The filtering of the time-series arrays}

Despite the fact that Refs.~\cite{winget1991,costa2008} read with pleasure, I am not sure that I grasp the application of the corrections to the raw data\footnote{Both papers refer to earlier work as to the processing of the 
raw measurements. I have not read these earlier papers.}. Some remarks on the data processing may be found on pp.~327--329 of Ref.~\cite{winget1991}: ``The effects of extinction and other slow transparency variations were 
accounted for by fitting a third-degree polynomial to each sky-subtracted data set and then dividing by this fit. This also normalised the data, so we then subtracted $1$ to give a mean of zero for all the data sets. This 
procedure yields a light curve with variation in amplitude as a function of the total intensity; such reduced light curves from different sites can then be combined without further processing. We used the data obtained on a 
nearby comparison star in channel $2$ to measure sky transparency. Data contaminated by cloud were discarded. The final product of these basic reduction procedures is shown in [their] Fig.~1, the light curve of the $6.5$ day 
interval when all the observatories were online. Note that the overlaps - where data were obtained simultaneously at two sites at different longitude - are visible only by the increased density of points; there are no 
discontinuities in the light curve that arose from `stitching' the runs from individual sites into the composite whole.'' On p.~629 of Ref.~\cite{costa2008}, one reads that the authors ``fitted a polynomial of $4$-th order to 
the light curve of each individual night, but even so, residual frequencies with considerable amplitude persisted in the residual light curve.'' The authors conclude: ``To eliminate them, we used a \emph{high-pass} filter, an 
algorithm that detects and eliminates signals with high amplitudes and frequencies lower than $300~\mu$Hz.''

In relation to the pulsation modes of PG 1159-035, Refs.~\cite{winget1991,costa2008} make it clear that the range of interest (in frequency) lies between $1$ and $3$ mHz. This may be taken to suggest that the contributions (to 
the variation of the luminosity of PG 1159-035) from frequencies outside this range do not originate from $g$-wave pulsations, but are due to other (uninteresting in the context of these works) phenomena. Evidently, one way to 
suppress the influence of such phenomena is by band-passing the original time series. The validity of this filtering procedure (e.g., the choice of the corner frequencies) rests upon our understanding of the physical processes 
underlying the observations of the physical system, namely of the way in which the ultra-dense matter behaves.

The following questions are relevant in case of application of a band-pass filter:
\begin{itemize}
\item Which is the best filtering method in the particular problem?
\item Which filtering order should be used?
\item Which are the band-pass corner frequencies?
\item Which are the band-stop corner frequencies?
\item How large should the band-pass ripple be?
\item How large should the band-stop attenuation be?
\end{itemize}
Although some of these questions can be answered by methods contained in my C/C++ library, it is more convenient (and often faster) to use MATLAB \textregistered~(The MathWorks, Inc., Natick, Massachusetts, United States) in 
filtering applications; I used MATLAB 7.5.0 in this work. Four filtering methods have been examined: Butterworth, Chebyshev (types I and II), and elliptic filters. In each case, the best filtering order was obtained from the 
desired band-pass/stop corner frequencies, the band-pass ripple, and the band-stop attenuation, using the MATLAB methods buttord, cheb1ord, cheb2ord, and ellipord, corresponding to the four aforementioned filtering methods, 
in that order. The application of the best filter was then enabled with the use of the MATLAB methods butter, cheby1, cheby2, and ellip, corresponding again to the four aforementioned filtering methods, in that order.

\begin{figure}
\begin{center}
\includegraphics [width=15.5cm] {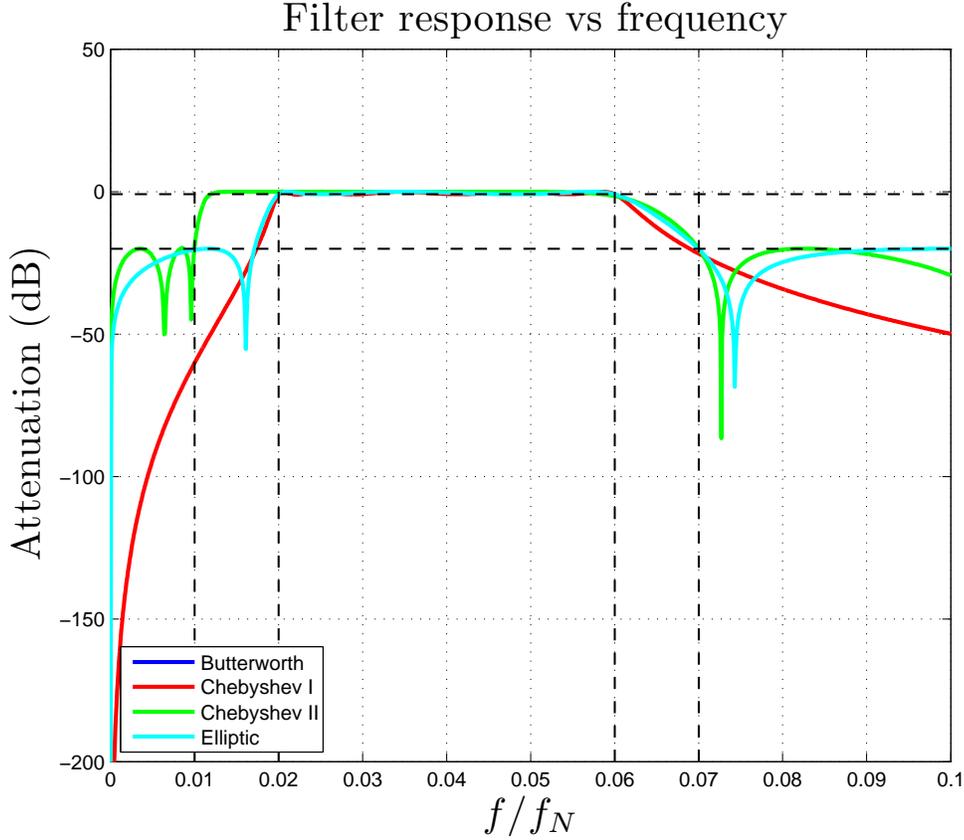}
\caption{\label{fig:ComparisonOfFilters}A comparison of the response functions obtained with the Chebyshev type I, the Chebyshev type II, and the elliptic filters for the filter properties given in Section \ref{sec:Filtering}. 
The Butterworth filtering method was also examined, but the results were not reasonable and are not shown in the figure. The Nyquist frequency $f_N$ for the measurements of PG 1159-035 is equal to $50$ mHz.}
\vspace{0.35cm}
\end{center}
\end{figure}

The left and right band-pass frequencies were set equal to $1$ and $3$ mHz, respectively. The left and right band-stop frequencies were set equal to $0.5$ and $3.5$ mHz, respectively. A few options were examined in relation to 
the band-pass ripple and the band-stop attenuation. By decreasing the former, one obtains filter response functions which vary less in the band-pass region; by increasing the latter, one obtains filter response functions which 
vary less in the band-stop region. Evidently, the ideal filter would result in a vanishing band-pass ripple and in complete band-stop attenuation. Of course, the ideal filter is a fictitious concept. In an effort to obtain 
filters which resemble better the ideal one, one may try to suppress the band-pass ripples and increase the band-stop attenuation, yet the order of the required filter increases the closer it comes to the ideal filter. In 
applications, this is not only time-consuming, but also prone to instability. Last but not least, high-order filters usually generate more delay between the original and the filtered data. Evidently, the selection of the best 
filter is a trade-off process between the ideal (application of Heaviside step functions) and the practical (stability, acceptable delay) filtering. The investigation suggested the use of the following parameters.
\begin{itemize}
\item Peak-to-peak band-pass ripple: $10 \%$, corresponding to about $-0.915$ dB.
\item Peak-to-peak band-stop attenuation: $90 \%$, corresponding to $-20$ dB.
\end{itemize}
(The unit dB stands for `decibel'.) The peak-to-peak band-pass ripple of $10 \%$ allows the filter response function to vary in the band-pass region between $90 \%$ and $100 \%$ of the peak value; the peak-to-peak band-stop 
attenuation of $90 \%$ forces the filter response function to remain below $10 \%$ of the peak value in the band-stop region. Even with these loose conditions, the Butterworth filter did not produce stable results. The results 
obtained with the remaining three filtering methods (see Fig.~\ref{fig:ComparisonOfFilters}) were found acceptable. The elliptic filter was finally chosen, as it required the lowest order ($6$-th), thus achieving the filtering 
of the time-series arrays with $13$ recursion coefficients. (For the Chebyshev type I and II filtering methods, $10$-th order filters were suggested, whereas the failing Butterworth method required a $22$-nd order filter.)

In Infinite Impulse Response (IIR) digital filtering, an $M$-order filter, applied (left-to-right) to the input time series, follows the recurrence relationship\footnote{The role of the recursion coefficients $a$ and $b$ 
frequently appears interchanged in the literature. I consider it more `natural' to apply the coefficients $a$ to the original data and the coefficients $b$ to the filtered ones.}:
\begin{equation} \label{eq:EQ003}
y_i = a_0 x_i + \sum_{k=1}^M \left( a_k x_{i-k} - b_k y_{i-k} \right) \, \, \,
\end{equation}
where $x_i$ stands for the input time-series array, $y_i$ for the filtered array. The constants $a_k$ and $b_k$ are known as recursion coefficients. (More general forms have appeared in the literature, e.g., allowing for 
different dimensions of the two arrays of recursion coefficients.) The coefficients of the elliptic filter, applied to the original time-series arrays of this work, are given in Table \ref{tab:RecursionCoefficients}.

\begin{table}
{\bf \caption{\label{tab:RecursionCoefficients}}}The recursion coefficients (see Eq.~(\ref{eq:EQ003})) of the elliptic filter used in band-passing the original time-series arrays of this work.
\vspace{0.2cm}
\begin{center}
\begin{tabular}{|c|c|}
\hline
Recursion coefficient & Value\\
\hline
\hline
$a_0$ & $1.936650845443 \cdot 10^{-2}$\\
$a_1$ & $-7.636547752052 \cdot 10^{-2}$\\
$a_2$ & $9.463412534230 \cdot 10^{-2}$\\
$a_3$ & $6.020320205958 \cdot 10^{-17}$\\
$a_4$ & $-9.463412534230 \cdot 10^{-2}$\\
$a_5$ & $7.636547752052 \cdot 10^{-2}$\\
$a_6$ & $-1.936650845443 \cdot 10^{-2}$\\
\hline
$b_1$ & $-5.827902199708 \cdot 10^{0}$\\
$b_2$ & $1.419838266997 \cdot 10^{1}$\\
$b_3$ & $-1.850916256654 \cdot 10^{1}$\\
$b_4$ & $1.361687656364 \cdot 10^{1}$\\
$b_5$ & $-5.360263930430 \cdot 10^{0}$\\
$b_6$ & $8.820710201218 \cdot 10^{-1}$\\
\hline
\end{tabular}
\end{center}
\vspace{0.5cm}
\end{table}

The seventeen time series of Ref.~\cite{timeseries} were submitted to this filtering procedure. The properties of the resulting time series are given in Table \ref{tab:FilteredData}. Many discrepancies, observed when processing 
the original data, have disappeared. It is comforting to see that the values of the embedding delay $\nu$ come out consistent after the filtering, namely between $12$ and $14$ sampling intervals in all cases. In short, the 
luminosity measurements of PG 1159-035 may be considered temporally uncorrelated if separated by slightly over $2$ minutes. Equally comforting is the assertion of the stationarity of the observations, as revealed by the p-values, 
obtained from the $\chi^2$ of Eq.~(\ref{eq:EQ001}) and the number of degrees of freedom $n-1$ in each data set; all p-values exceed the significance threshold $\mathrm{p}_{\rm min}$ of this work ($1.00 \cdot 10^{-2}$).

\begin{table}
{\bf \caption{\label{tab:FilteredData}}}Some quantities relating to the filtered time-series arrays of PG 1159-035. The filtering of the original measurements is discussed in Section \ref{sec:Filtering}.
\vspace{0.2cm}
\begin{center}
\begin{tabular}{|c|c|c|c|c|c|c|c|}
\hline
Data set & $s_{\rm min}$ & $s_{\rm max}$ & $s_{\rm max}-s_{\rm min}$ & $\avg{s}$ & rms & p-value & $\nu$\\
\hline
\hline
E01 & $-0.2419$ & $0.2425$ & $0.4844$ & $3.026 \cdot 10^{-4}$ & $1.156 \cdot 10^{-1}$ & $9.82 \cdot 10^{-1}$ & $14$\\
E02 & $-0.1957$ & $0.2162$ & $0.4119$ & $-4.645 \cdot 10^{-5}$ & $8.181 \cdot 10^{-2}$ & $9.24 \cdot 10^{-1}$ & $13$\\
E03 & $-0.1624$ & $0.1493$ & $0.3117$ & $1.018 \cdot 10^{-4}$ & $7.037 \cdot 10^{-2}$ & $8.07 \cdot 10^{-1}$ & $13$\\
E04 & $-0.1401$ & $0.1680$ & $0.3081$ & $1.210 \cdot 10^{-4}$ & $7.372 \cdot 10^{-2}$ & $4.95 \cdot 10^{-1}$ & $14$\\
E05 & $-0.1257$ & $0.1117$ & $0.2374$ & $-1.590 \cdot 10^{-4}$ & $5.111 \cdot 10^{-2}$ & $7.14 \cdot 10^{-1}$ & $12$\\
E06 & $-0.1822$ & $0.2029$ & $0.3850$ & $1.125 \cdot 10^{-4}$ & $7.212 \cdot 10^{-2}$ & $6.50 \cdot 10^{-1}$ & $13$\\
E07 & $-0.2279$ & $0.2147$ & $0.4426$ & $4.551 \cdot 10^{-5}$ & $8.888 \cdot 10^{-2}$ & $5.39 \cdot 10^{-2}$ & $14$\\
E08 & $-0.3259$ & $0.3286$ & $0.6544$ & $-7.094 \cdot 10^{-5}$ & $1.206 \cdot 10^{-1}$ & $6.59 \cdot 10^{-1}$ & $13$\\
E09 & $-0.3340$ & $0.3340$ & $0.6680$ & $5.784 \cdot 10^{-5}$ & $1.141 \cdot 10^{-1}$ & $3.12 \cdot 10^{-1}$ & $14$\\
E10 & $-0.3270$ & $0.3083$ & $0.6352$ & $6.431 \cdot 10^{-5}$ & $1.179 \cdot 10^{-1}$ & $4.29 \cdot 10^{-1}$ & $14$\\
E11 & $-0.2179$ & $0.2829$ & $0.5008$ & $1.190 \cdot 10^{-2}$ & $9.538 \cdot 10^{-2}$ & $4.98 \cdot 10^{-2}$ & $14$\\
E12 & $-0.2609$ & $0.2379$ & $0.4988$ & $6.363 \cdot 10^{-5}$ & $8.584 \cdot 10^{-2}$ & $4.23 \cdot 10^{-1}$ & $13$\\
E13 & $-0.2784$ & $0.2716$ & $0.5499$ & $-5.841 \cdot 10^{-7}$ & $9.428 \cdot 10^{-2}$ & $1.24 \cdot 10^{-1}$ & $13$\\
E14 & $-0.2750$ & $0.2583$ & $0.5333$ & $-1.653 \cdot 10^{-3}$ & $9.867 \cdot 10^{-2}$ & $5.20 \cdot 10^{-2}$ & $13$\\
E15 & $-0.2979$ & $0.2112$ & $0.5090$ & $-3.384 \cdot 10^{-4}$ & $9.970 \cdot 10^{-2}$ & $6.16 \cdot 10^{-1}$ & $14$\\
E16 & $-0.2222$ & $0.2414$ & $0.4636$ & $5.221 \cdot 10^{-3}$ & $8.770 \cdot 10^{-2}$ & $5.11 \cdot 10^{-1}$ & $13$\\
E17 & $-0.1640$ & $0.1686$ & $0.3327$ & $1.327 \cdot 10^{-4}$ & $6.247 \cdot 10^{-2}$ & $3.05 \cdot 10^{-1}$ & $13$\\
\hline
\end{tabular}
\end{center}
\vspace{0.5cm}
\end{table}

\section{\label{sec:AnalysisFiltered}The analysis of the filtered time-series arrays}

\subsection{\label{sec:Database}Creation of a database of embedding vectors}

To determine whether the physical processes, underlying the observations, are linear or non-linear, one needs to perform a non-linear analysis of the measurements. Each element of a time series may be considered as an instance 
of the `present'; the preceding elements may be thought of as representing the element's `past', whereas subsequent ones its `future'. Therefore, in a given time series, the past and the future of each `snapshot of the present' 
are known (of course, the extent to which the past and the future of an element are known depends on the position of that element in the time series), save for the first element (whose past is unknown) and the last one (whose 
future is unknown). The determination of whether a system is linear or non-linear rests upon the extraction of predictions from the known past of each element of the time series and their subsequent comparison to the 
measurements representing that element's future. To this end, a database of embedding vectors was created, along with the associated predictions for $\Delta n=100$ sampling intervals into the future of the last element of each 
such vector. A time-series array containing $N$ elements yields $N - \Delta n - (m-1) \nu$ embedding vectors. The predictions, obtained from the database of embedding vectors, will be used for estimating the out-of-sample 
prediction errors.

The data sets were split into two parts: one part yielded the training (or learning) set, from which the database embedding vectors were obtained, the other the test set, which provided the vectors, whose neighbours are sought 
among the database entries. These two independent sets were created as follows. The seventeen data sets of Table \ref{tab:FilteredData} were first ordered in terms of the number of measurements they contain. The data set with 
the largest content (i.e., the data set E11) was assigned to the training set. Each subsequent pair of files were assigned either to the test or to the training sets alternatively (starting with the test set) until seven data 
sets had been selected for the training set. The last four files were assigned to the test set. Selected for the training set were the data sets: E11, E08, E09, E06, E12, E02, and E03. Selected for the test set were the data 
sets: E14, E13, E07, E17, E16, E10, E04, E15, E01, and E05. As a result, $12617$ measurements in total were assigned to the training set, $14574$ to the test set. The distributions of the minimal and maximal signals, of the 
range, of the average, and of the rms values between the two sets were subjected to tests for significant differences (two-tailed, homoscedastic t-tests); none were found.

Given that the seventeen data sets represent luminosity measurements of a stationary (during the temporal span of the observations) process, it makes sense to use one lag value $\nu$ in the analysis. This choice is not a matter 
of convenience, but one of rationality. (In fact, the implementation has been made in such a way that it covers the general case, i.e., variable lag.) The maximal lag from Table \ref{tab:FilteredData}, namely $\nu = 14$, will 
be used in the remaining part of this paper. As any two luminosity measurements are assumed independent if temporally separated by $\nu$ sampling intervals, the choice of the maximal lag value in Table \ref{tab:FilteredData} 
ensures that the components of all embedding vectors, be they related to the training or to the test set, will be independent.

\subsection{\label{sec:Distance}Distance in an $m$-dimensional space}

Before advancing, one word about the definition of the distance $d$ between two $m$-dimensional vectors $a_k$ and $b_k$ is due. The general definition of the distance is:
\begin{equation*}
d = \left( \sum_{k=1}^m \abs{ a_k - b_k }^n \right)^{1/n} \, \, \, ,
\end{equation*}
where $n=\{ 1, \dots, +\infty \}$ sets the norm. The Euclidean distance corresponds to $n=2$ ($L^2$ distance). The $L^1$-norm distance is the sum of the absolute values of the differences between the components of the two vectors
\begin{equation*}
d = \sum_{k=1}^m \abs{ a_k - b_k } \, \, \, ,
\end{equation*}
whereas the $L^\infty$-norm distance satisfies
\begin{equation} \label{eq:EQ003_5}
d = {\rm max} \{ \abs{ a_k - b_k } \} _{k=1}^m \, \, \, .
\end{equation}

Although the main results of this work will be obtained with the $L^\infty$ distance, use of the $L^2$ distance will also be made, as a verification step, in Section \ref{sec:L2Norm}.

\subsection{\label{sec:EmbeddingDimension}On the optimal embedding dimension and range of variation of the neighbourhood size}

As mentioned in Section \ref{sec:AnalysisOriginal}, the determination of the optimal embedding dimension is an important step in phase-space reconstructions. If the dimensionality of the phase space is $D$, Takens' theorem 
\cite{takens1981} ensures that embeddings exist in a mathematical space of more than $2 D$ dimensions, which fully uncover the characteristics of the phase space. Takens' theorem does not \emph{prohibit} sufficient embeddings 
in fewer dimensions\footnote{To mention two examples: the correlation dimension (to be introduced in Section \ref{sec:CorrelDim}) for the attractor of the H{\'e}non map has been estimated to $1.25 \pm 0.02$ \cite{grassberger1983}, 
yet the optimal embedding dimension is $2$ \cite{cao1997}; for the Lorenz attractor, the correlation dimension has been estimated to $2.05 \pm 0.01$ \cite{grassberger1983} and the optimal embedding dimension is $3$ \cite{cao1997}.}; 
it simply does not \emph{guarantee} their existence. In any case, the minimal embedding dimension, leading to a sufficient embedding, will be called optimal from now on (and will be denoted by $m_0$). Obviously, the use of 
embedding dimensions below the optimal one yields embeddings which are insufficient: they cannot cover the phase space.

As when addressing the optimal embedding delay in Section \ref{sec:AnalysisOriginal}, one question arises: if the objective is to obtain sufficient embeddings, why should one not choose an embedding dimension so large that 
the probability of an insufficient embedding be practically zero? The answer is that large embedding dimensions introduce complexity and redundancy, and hinder the interpretation of the results of an analysis. KS give two 
additional arguments: large embedding dimensions demand large computational effort (which became less of a problem during the last two decades) and (more importantly) they lead to the degradation of the performance of the 
algorithms used in non-linear analysis.

A number of methods have been put forward to provide a reliable estimate for the optimal embedding dimension. One category of methods pivot on the technique of the `false nearest neighbours' (FNN), introduced by Kennel, Brown, 
and Abarbanel in 1992 \cite{kennel1992}. The idea behind the method is quite simple. As $m_0$ is the optimal embedding dimension, the set of embedding vectors, neighbouring an arbitrary vector, is expected to be left intact in 
embeddings in $m_0+1$ dimensions. In case of an insufficient embedding (i.e., when $m<m_0$), all embedding vectors, corresponding to the true neighbours in the $m_0$-dimensional embedding, will also be contained in the set of 
neighbours in the $m$-dimensional embedding; this is guaranteed by the definition of the neighbourhood. However, given that the $m$-dimensional embedding is not the optimal one, $m_0$-dimensional vectors exist which, though 
not belonging to the neighbourhood in $m_0$ dimensions, are projected unto neighbouring vectors in the $m$-dimensional embedding. These are the false neighbours, i.e., states which are ostensibly neighbouring (a specific state) 
only because the embedding dimension is insufficient.

Several variants of the FNN method have appeared in the literature, e.g., see Refs.~\cite{kantz1997,cao1997,krakovska2015} (this list is anything but exhaustive). Cao's method \cite{cao1997} is my favourite for three reasons: 
a) it is a parameter-free approach (save for the lag), b) it is straightforward to implement, and c) it is robust. In Section \ref{sec:Cao}, I will obtain an estimate for the optimal embedding dimension of the luminosity 
measurements of PG 1159-035 using this method. Another category of methods involve the $m$-dependence of a dynamical invariant, i.e., of a quantity characterising the phase space, e.g., of the correlation dimension or of the 
maximal Lyapunov exponent. Interesting overviews on this subject may be found in Refs.~\cite{cao1997,krakovska2015}, as well as in the works cited therein.

The choice for the neighbourhood size $\epsilon$, entering the determination of the correlation dimension and the extraction of the maximal Lyapunov exponent, is addressed in Refs.~\cite{kantz1997}. As there is no concrete, 
theoretically justified way to select $\epsilon$, the determination of the appropriate range of values remains, to a large extent, empirical; frequently, the trial-and-error method is the only practical approach. KS write: 
``Studies with known true signals suggest that a good choice for the neighbourhood size is given by $2-3$ times the noise amplitude.'' However, this recommendation cannot easily be followed if a noise-reduction scheme has 
been applied to the input data. As a remedy against the lack of a concrete theoretical basis for the determination of the $\epsilon$ domain, KS suggest the use of several $\epsilon$ values, and frequently in their book they 
vary $\epsilon$ within a range covering a few orders of magnitude. Evidently, $\epsilon$ may be thought of as a free parameter in a non-linear analysis: KS recommend that the only prerequisite in the variation of $\epsilon$ 
be the existence of an adequate number of neighbours within the $m$-dimensional balls (or boxes) corresponding to the different embeddings.

The average rms of the measurements of PG 1159-035 (filtered time-series arrays) is equal to about $9.00 \cdot 10^{-2}$, see Table \ref{tab:FilteredData}. It was decided to use half the average rms as the starting value of 
$\epsilon$, and decrease this parameter (using a small step) until the number of neighbours dropped to such a level that the statistical analysis was possible only for low-dimensional embeddings (e.g., below $6$).

\subsection{\label{sec:Cao}Cao's method for the determination of the optimal embedding dimension}

In the abstract of his 1997 paper, Cao lists the advantages of his method for determining the optimal embedding dimension: the method ``(1) does not contain any subjective parameters except for the time-delay for the embedding; 
(2) does not strongly depend on how many data points are available; (3) can clearly distinguish deterministic signals from stochastic signals; (4) works well for time series from high-dimensional attractors; (5) is 
computationally efficient.''

Using the $L^\infty$ distance, Cao's method examines the dependence of two quantities, named $E1$ and $E2$ in the paper (see Eqs.~($1-5$) in Ref.~\cite{cao1997}), on the embedding dimension: $E1$ represents the relative change 
of the average distance between neighbouring embedding vectors when increasing the embedding dimension by one unit, whereas $E2$ essentially tests the independence of past and future values. As Cao remarks, his method 
distinguishes deterministic and stochastic signals: in the latter case, $E2$ comes out close to $1$ regardless of the embedding dimension. For deterministic signals, $E1$ and $E2$ approach saturation with increasing embedding 
dimension: the optimal embedding dimension is chosen to be the one at which $E1$ saturates, i.e., it does not change (significantly) when further increasing $m$. Cao recommends the evaluation and visual inspection of $E2$, as 
the means to ensure that the signal analysed is indeed deterministic.

\begin{figure}
\begin{center}
\includegraphics [width=15.5cm] {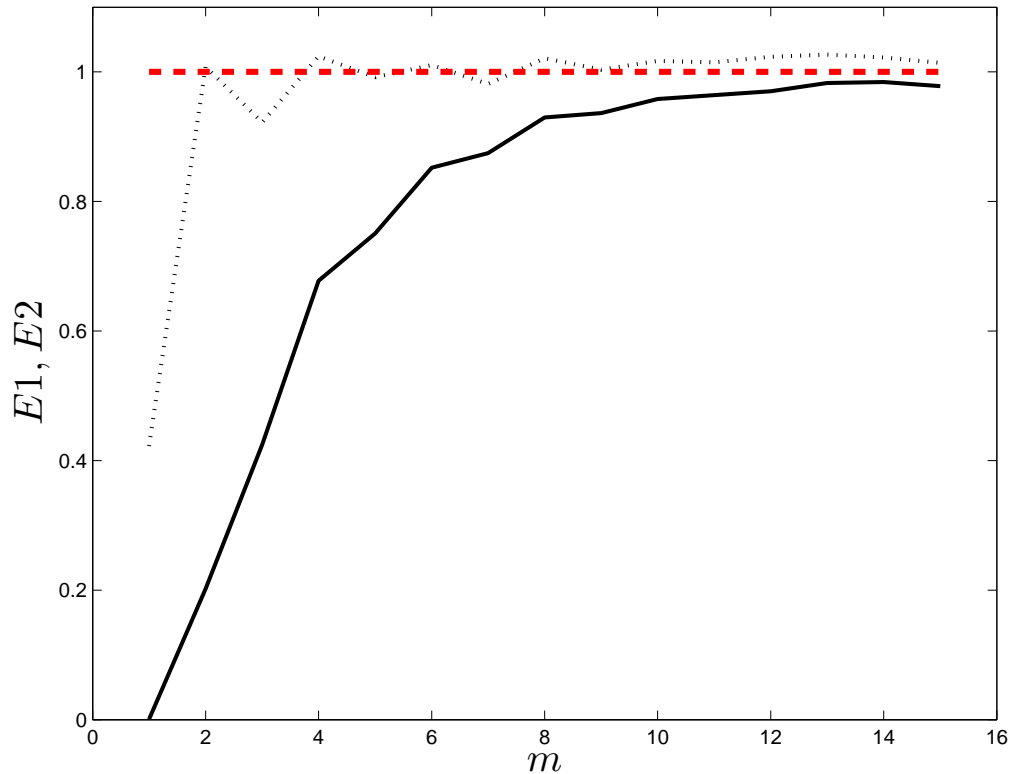}
\caption{\label{fig:CaoFromDatabase}Cao's $E1$ (straight line segments) and $E2$ (dotted line segments) for embedding dimensions $m$ up to $15$. The training and test sets, as defined in Section \ref{sec:AnalysisFiltered}, 
have been used.}
\vspace{0.35cm}
\end{center}
\end{figure}

The quantities $E1$ and $E2$ were evaluated for embedding dimensions up to $m=15$, using (as the only input) $\nu=14$, and the training and test sets, as defined in Section \ref{sec:AnalysisFiltered}. The results are shown 
in Fig.~\ref{fig:CaoFromDatabase}.

\begin{figure}
\begin{center}
\includegraphics [width=15.5cm] {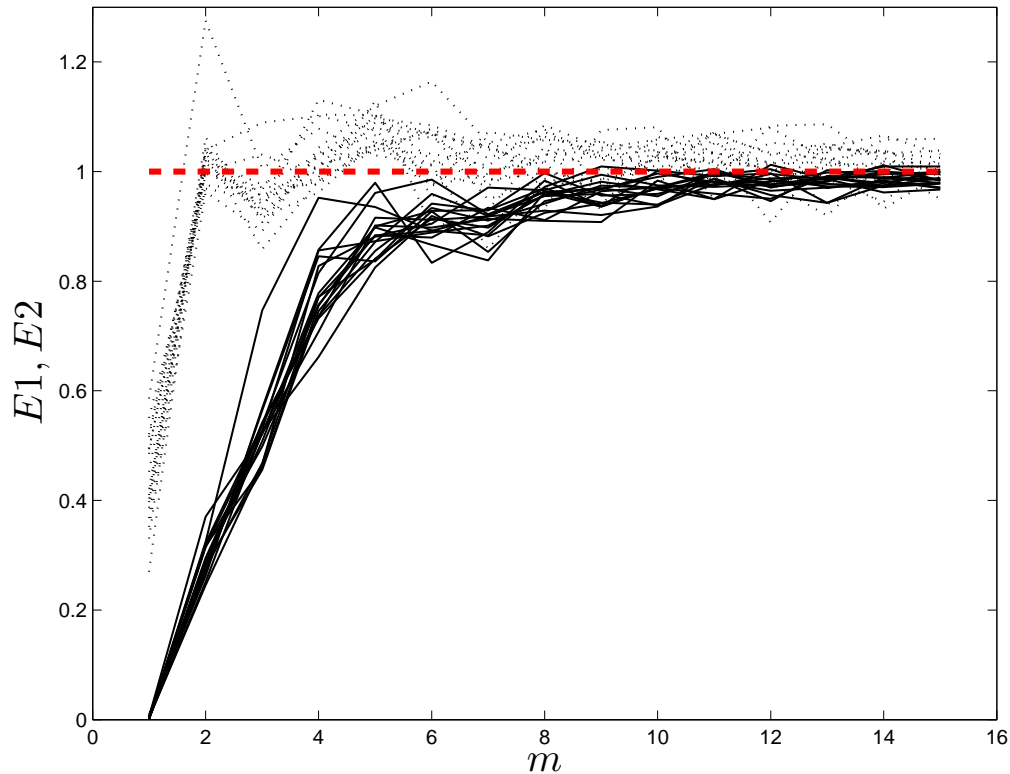}
\caption{\label{fig:Cao}Cao's $E1$ (straight line segments) and $E2$ (dotted line segments) for embedding dimensions $m$ up to $15$. These quantities have been obtained from a separate analysis of the seventeen data sets (no 
splitting of the data into training and test sets). All contributing embedding vectors were required to have a temporal separation exceeding $\nu \Delta \tau$.}
\vspace{0.35cm}
\end{center}
\end{figure}

Although one extracts a large estimate for the optimal embedding dimension from Fig.~\ref{fig:CaoFromDatabase}, perhaps around $13$, it might make sense to accept that $E1$ already saturates in the vicinity of $10$. This choice 
is confirmed by the result of the separate analysis\footnote{To reduce the temporal correlations in the separate analysis of the data sets, all contributing (to the determination of $E1$ and $E2$) embedding vectors were required 
to have a temporal separation (constant distance between their corresponding elements) exceeding the embedding delay $\nu$. Temporal separations of $2 \nu \Delta \tau$ and $3 \nu \Delta \tau$ have also been attempted, but 
induced very small differences on the results.} of the seventeen input data sets, shown in Fig.~\ref{fig:Cao}. There is general agreement between Figs.~\ref{fig:CaoFromDatabase} and \ref{fig:Cao}: from these two figures, it 
appears reasonable to accept the value of $10$ as the optimal embedding dimension.

\subsection{\label{sec:CorrelDim}Correlation dimension}

The notion of the correlation dimension was introduced by Grassberger and Procaccia in 1983 \cite{grassberger1983}: it is a measure of the dimensionality of the phase space. Let us examine how the correlation dimension is 
obtained from the correlation sum, which represents the frequentness of embedding vectors in the time series whose distance is below a given neighbourhood size $\epsilon$:
\begin{equation} \label{eq:EQ004}
C(\epsilon) = \frac{2}{N (N-1)} \sum_{i=1}^{N-1} \sum_{j=i+1}^{N} \Theta (\epsilon - \norm {{\ss}_i - {\ss}_j}) \, \, \, ,
\end{equation}
where $\Theta(x)$ is the Heaviside step function, attaining the values of $0$ when $x<0$, and $1$ when $x \geq 0$. To suppress the temporal correlations, KS propose a modification of Eq.~(\ref{eq:EQ004}):
\begin{equation} \label{eq:EQ005}
C(\epsilon) = \frac{2}{(N-n_{\rm min}) (N-1-n_{\rm min})} \sum_{i=1}^{N-1-n_{\rm min}} \sum_{j=i+1+n_{\rm min}}^{N} \Theta (\epsilon - \norm {{\ss}_i - {\ss}_j}) \, \, \, ,
\end{equation}
and recommend the use of a generous $n_{\rm min}$, significantly exceeding the lag extracted from the autocorrelation function. (The expressions, given by KS in both editions of their book, need to be corrected: the upper 
limits of the first sums are misleading. In addition, one unit must be added to the lower limit of the second sum of the equation featuring $n_{\rm min}$ in the first edition. Equations (\ref{eq:EQ004},\ref{eq:EQ005}), as 
they appear in this work, are the correct expressions.)

It is expected that, in the limit $N \to \infty$, $C(\epsilon)$ would scale like a power law at small $\epsilon$ values, i.e., $C(\epsilon) \sim \epsilon^D$. The correlation dimension $D$ would thus be obtained as a double limit.
\begin{equation} \label{eq:EQ005_5}
\alpha (N,\epsilon)=\frac{\partial \ln C(\epsilon)}{\partial \ln \epsilon}
\end{equation}
\begin{equation*}
D=\lim_{N \to \infty} \lim_{\epsilon \to 0} \alpha (N,\epsilon)
\end{equation*}
The practical application of these notions to a time series involves the stability of $\alpha (N,\epsilon)$ when plotted as a function of $\epsilon$ for several embedding dimensions. KS remark that the estimation of the 
correlation dimension should be thought of as a two-step process. In the first step, the correlation sums are evaluated for several $\epsilon$ and $m$ values. The second step involves the visual inspection of the 
($\ln \epsilon$,$\ln C(\epsilon)$) scatter plots: if an $\epsilon$ domain exists, within which the dependence of $\ln C(\epsilon)$ on $\ln \epsilon$ is linear for all sufficient embedding dimensions, then one may have 
confidence in the extracted dimensionality of the phase space. In this work, estimates for $\alpha$ will be obtained directly from linear fits to the correlation sums.

Equations (\ref{eq:EQ004},\ref{eq:EQ005}) pertain to the extraction of the correlation sums from one input time-series array: in that case, no database is available and the source of the embedding vectors must inevitably be 
the time-series array itself. The modification is straightforward if the division of the measurements into training and test sets is an option, as the case is in this work.

Let the database contain $M(m)$ embedding vectors $\tilde{\ss}_i$ for an embedding dimension $m$. Let also $N_k (m)$ be the number of embedding vectors, which can be constructed from the $k$-th test set, and $N_t$ the number 
of files in the test set. The correlation sum for embedding dimension $m$ would then be defined as
\begin{equation} \label{eq:EQ006}
C(\epsilon) = \left( M(m) \sum_{k=1}^{N_t} N_k (m) \right)^{-1} \sum_{j=1}^{M(m)} \sum_{k=1}^{N_t} \sum_{i=1}^{N_k (m)} \Theta (\epsilon - \norm {{\ss}_{i,k} - \tilde{\ss}_j}) \, \, \, ,
\end{equation}
where ${\ss}_{i,k}$ denotes the $i$-th embedding vector of the $k$-th test set.

To obtain reliable estimates for the correlation dimension, the $C(\epsilon)$ arrays were not processed if the triple sum in Eq.~(\ref{eq:EQ006}) yielded fewer than $10$ (non-zero) contributions. The correlation sums were 
obtained at $40$ $\epsilon$ values (i.e., between $4.50 \cdot 10^{-2}$ and $0.60 \cdot 10^{-2}$, with a step of $0.10 \cdot 10^{-2}$) and $10$ embedding dimensions (i.e., $3 \leq m \leq 12$). The reason for choosing such a 
wide domain of embedding dimensions, in spite of having obtained $m_0$ in Section \ref{sec:Cao}, is simple: this choice enables the study of the behaviour of the estimates for the dynamical invariants, e.g., for the correlation 
dimension and for the maximal Lyapunov exponent, in the transition from insufficient embedding dimensions to sufficient ones.

\begin{figure}
\begin{center}
\includegraphics [width=15.5cm] {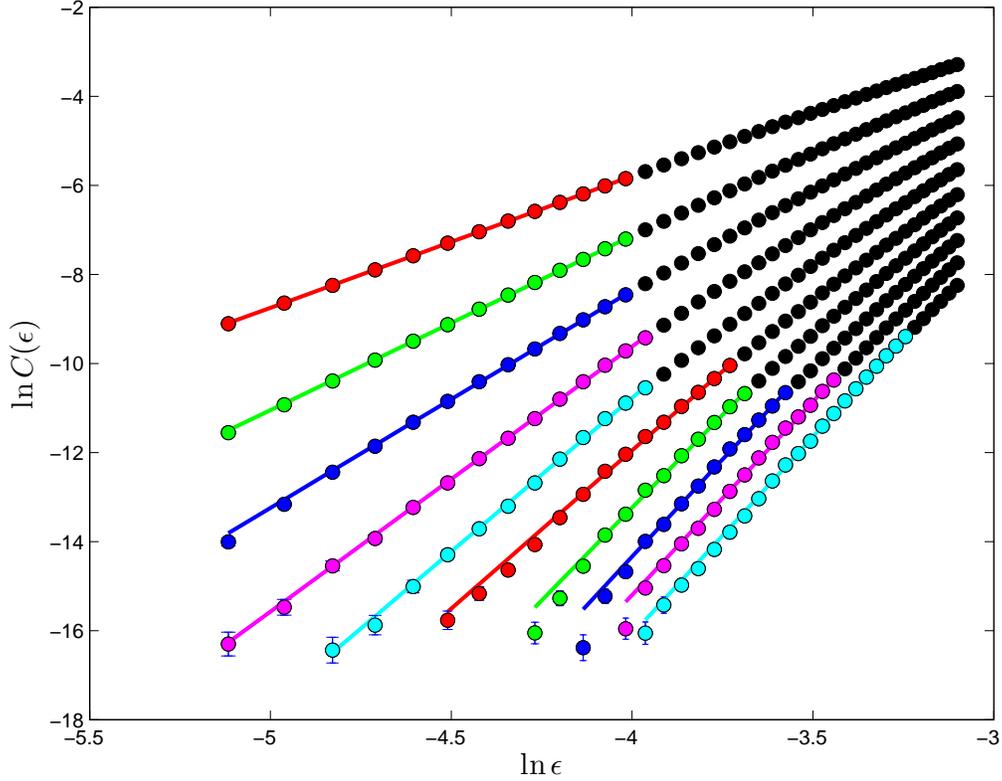}
\caption{\label{fig:CorrelationSums}The ($\ln \epsilon$,$\ln C(\epsilon)$) scatter plots for embedding dimensions $m=3$ (top) to $m=12$ (bottom). Weighted least-squares fits were performed on the data (see Table \ref{tab:DimensionLinf}), 
separately for each embedding dimension in the $\epsilon$ domain within which the linearity between $\ln C(\epsilon)$ and $\ln \epsilon$ holds (coloured points and straight lines); the data outside these $\epsilon$ domains are 
also shown (in black). Although $\ln C(\epsilon)$ appears to depend linearly on $\ln \epsilon$ in broader $\epsilon$ domains, the rigorous test of linearity, based on the resulting $\chi^2$ value and the number of degrees of 
freedom in each fit, fails for $\epsilon$ domains broader than those shown in this figure.}
\vspace{0.35cm}
\end{center}
\end{figure}

\begin{figure}
\begin{center}
\includegraphics [width=15.5cm] {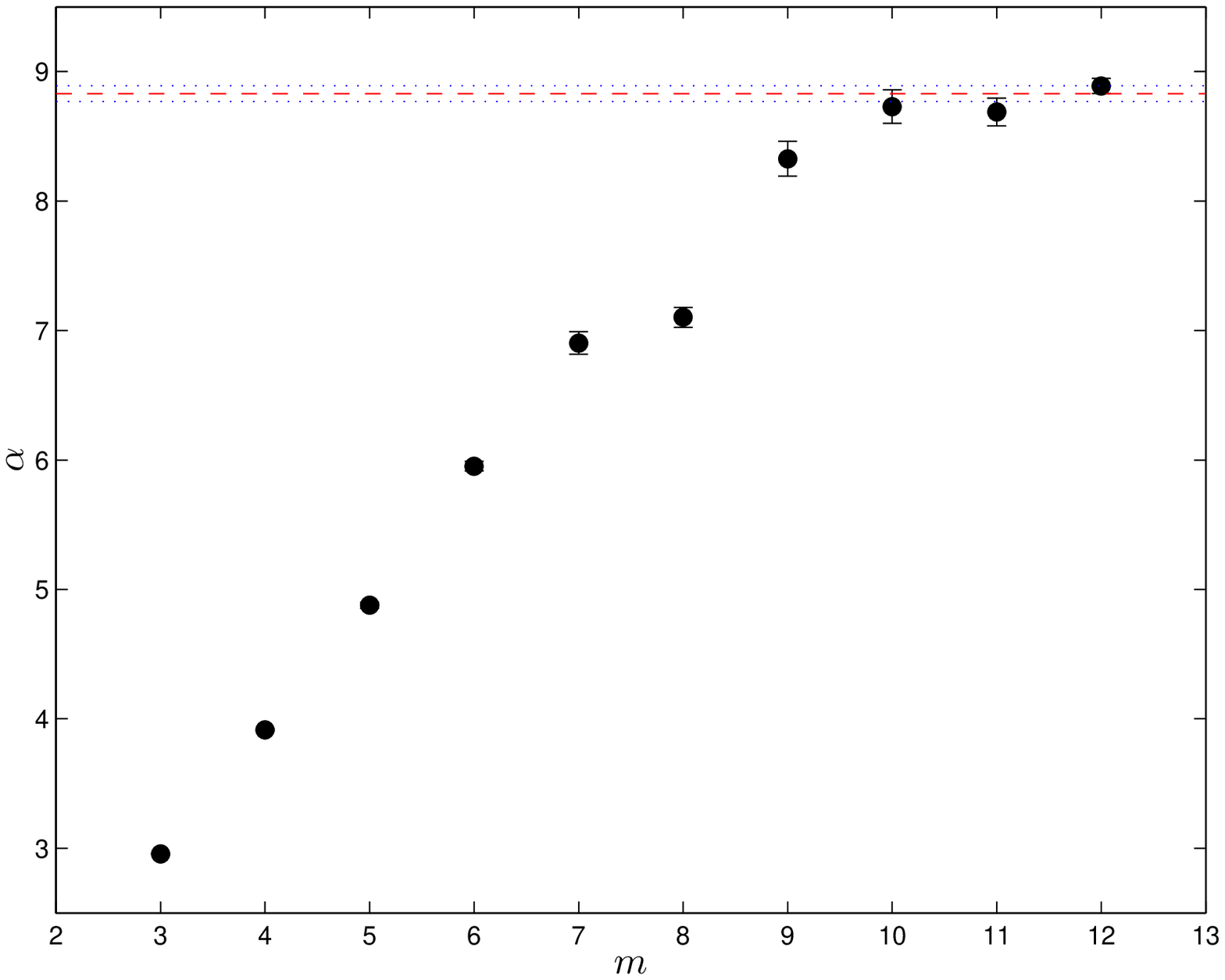}
\caption{\label{fig:CorrelationDimension}The slope $\alpha$ of the linear fit to the correlation sum $\ln C(\epsilon)$, see Eq.~(\ref{eq:EQ006_5}). The red dashed line represents the weighted average over the $\alpha$ values 
for the sufficient embeddings ($10 \leq m \leq 12$), see Table \ref{tab:DimensionLinf}. The blue dotted lines represent the $1 \sigma$ limits of the statistical uncertainty, corrected for the quality of the reproduction of the 
three input $\alpha$ values by their weighted average.}
\vspace{0.35cm}
\end{center}
\end{figure}

As explained earlier, one is interested in the domain of $\epsilon$ values for which the relationship between $\ln C(\epsilon)$ and $\ln \epsilon$ is linear; within this region,
\begin{equation} \label{eq:EQ006_5}
\ln C(\epsilon) = \alpha \ln \epsilon + \beta \, \, \, ,
\end{equation}
where the slope $\alpha$ is identified with $\alpha (N,\epsilon)$ of Eq.~(\ref{eq:EQ005_5}). At fixed $m$, the linearity between $\ln C(\epsilon)$ and $\ln \epsilon$ was investigated\footnote{References \cite{kantz1997} give 
the reader the impression that KS rather favour the visual inspection of the ($\ln \epsilon$,$\ln C(\epsilon)$) scatter plots as the means to establish the $\epsilon$ domain within which the linearity holds. My preference is 
to apply the appropriate statistical rules, and, if unable to perform an analysis as a result of the rigorousness of the conditions imposed, to rather relax these rigorous conditions in a consistent manner, e.g., by decreasing 
the threshold $\mathrm{p}_{\rm min}$ by one or two orders of magnitude.}, starting from the original ($\ln \epsilon$,$\ln C(\epsilon)$) points and removing the point with the largest $\epsilon$ value (one point per iteration), 
until the resulting p-value (obtained from the $\chi^2$ value and the number of degrees of freedom in the linear fit) exceeded $\mathrm{p}_{\rm min}$, the threshold of statistical significance (see Section \ref{sec:AnalysisOriginal}). 
Table \ref{tab:DimensionLinf}, and Figs.~\ref{fig:CorrelationSums} and \ref{fig:CorrelationDimension} contain the main results of the analysis of the correlation dimension for the problem dealt with in this paper.

\begin{table}
{\bf \caption{\label{tab:DimensionLinf}}}Results of the analysis of the correlation sums obtained using Eq.~(\ref{eq:EQ006}) for embedding dimensions between $3$ and $12$. The given domain $[ \epsilon_{\rm min}, \epsilon_{\rm max} ]$ 
corresponds to the $\epsilon$ domain within which the linearity between $\ln C(\epsilon)$ and $\ln \epsilon$ is accepted ($\mathrm{p} \geq \mathrm{p}_{\rm min}$).
\vspace{0.2cm}
\begin{center}
\begin{tabular}{|c|c|c|c|c|c|c|}
\hline
$m$ & $\epsilon_{\rm min}$ & $\epsilon_{\rm max}$ & $\alpha$ & $\delta \alpha$ & $\beta$ & $\delta \beta$\\
\hline
\hline
$3$ & $6.00 \cdot 10^{-3}$ & $1.80 \cdot 10^{-2}$ & $2.9573$ & $0.0041$ & $6.035$ & $0.017$\\
$4$ & $6.00 \cdot 10^{-3}$ & $1.80 \cdot 10^{-2}$ & $3.9156$ & $0.0097$ & $8.528$ & $0.041$\\
$5$ & $6.00 \cdot 10^{-3}$ & $1.80 \cdot 10^{-2}$ & $4.878$ & $0.023$ & $11.145$ & $0.094$\\
$6$ & $6.00 \cdot 10^{-3}$ & $1.90 \cdot 10^{-2}$ & $5.951$ & $0.037$ & $14.18$ & $0.15$\\
$7$ & $8.00 \cdot 10^{-3}$ & $1.90 \cdot 10^{-2}$ & $6.904$ & $0.088$ & $16.84$ & $0.36$\\
$8$ & $1.10 \cdot 10^{-2}$ & $2.40 \cdot 10^{-2}$ & $7.102$ & $0.077$ & $16.45$ & $0.30$\\
$9$ & $1.40 \cdot 10^{-2}$ & $2.50 \cdot 10^{-2}$ & $8.33$ & $0.14$ & $20.07$ & $0.51$\\
$10$ & $1.60 \cdot 10^{-2}$ & $2.80 \cdot 10^{-2}$ & $8.73$ & $0.13$ & $20.58$ & $0.48$\\
$11$ & $1.80 \cdot 10^{-2}$ & $3.20 \cdot 10^{-2}$ & $8.69$ & $0.11$ & $19.55$ & $0.38$\\
$12$ & $1.90 \cdot 10^{-2}$ & $3.90 \cdot 10^{-2}$ & $8.890$ & $0.057$ & $19.47$ & $0.19$\\
\hline
\end{tabular}
\end{center}
\vspace{0.5cm}
\end{table}

The slope $\alpha$ of the linear fit of Eq.~(\ref{eq:EQ006_5}) should not be $m$-dependent for all sufficient embedding dimensions; this is the case for $10 \leq m \leq 12$. This result confirms the choice of $10$ as the 
optimal embedding dimension in Section \ref{sec:Cao}. Assuming only sufficient embeddings ($10 \leq m \leq 12$), one obtains for $\alpha$ the weighted average of $8.830 \pm 0.062$.

\subsection{\label{sec:Lyapunov}Lyapunov exponents}

I will next address the extraction of the maximal Lyapunov exponent from the measurements. Many regard the inability to reliably predict the future of chaotic systems, in spite of the known past and of a deterministic evolution, 
as the prominent characteristic of the chaotic behaviour. Of course, one needs to quantify what is meant by `reliably' in the previous sentence. Predictions are routinely made in all systems, ordinary or chaotic (e.g., using 
autoregressive-moving-average (ARMA) models in linear analyses, Lorenz's method of analogues in non-linear ones). The matter is that, in chaotic systems, neither do these predictions scatter around the observations (whenever 
they become available), nor does the difference between predicted and measured values increase linearly with time: the differences in chaotic systems (between predicted and measured values) grow exponentially with time (up to 
the time when the distance between the embedding vectors saturates). The Lyapunov exponents (usually denoted by $\lambda$) characterise the rapidity of the exponential divergence between predictions and observations (before 
the saturation effects prevail). In fact, one is predominantly interested in the maximal Lyapunov exponent (from now on, $\lambda$ will stand for this exponent), as this quantity reflects the long-term behaviour of a chaotic 
system. The signature of deterministic chaos is a positive and finite $\lambda$ value ($0 < \lambda < \infty$).

KS recommend one procedure for the determination of $\lambda$ from a time-series array (of dimension $N$) and a database of embedding vectors (if no database is available, one may use the time-series array itself as the source 
of embedding vectors and predictions). The steps of this procedure may be summarised as follows.
\begin{itemize}
\item Fix the embedding dimension $m$ and the neighbourhood size $\epsilon$.
\item Commence with the element of the time-series array with index $i = 1 + (m-1) \nu$ (the index of the first element of the time-series array is assumed to be $1$).
\item Construct the embedding vector ${\ss}_i$ according to Eq.~(\ref{eq:EQ002}).
\item Search the database for embedding vectors $\tilde{\ss}_j$, satisfying
\begin{equation} \label{eq:EQ007}
d_{ij} \coloneqq \norm{ {\ss}_i - \tilde{\ss}_j } \leq \epsilon \, \, \, .
\end{equation}
Let the set of acceptable indices $j$ (of the embedding vectors in the database, which neighbour the vector ${\ss}_i$ in the context of Eq.~(\ref{eq:EQ007})) be denoted as $\mathscr{U}_i$.
\item Use the known futures in the original time series $s$ ($\Delta n$ sampling intervals after the $i$-th element) and in the database $\tilde{s}$ ($\Delta n$ sampling intervals after the last element of each embedding 
vector $\tilde{\ss}_j$) and store the absolute differences in an array:
\begin{equation} \label{eq:EQ008}
Q_i (k) = \frac{1}{N_b} \sum_{j \in \mathscr{U}_i} \abs{ s_{ i + k } - \tilde{s}_{ j + k } } \, \, \, ,
\end{equation}
for $k \in \{ 0, 1, \dots, \Delta n \}$; $N_b$ denotes the number of embedding vectors in $\mathscr{U}_i$.
\item Obtain the natural logarithm of $Q_i (k)$ and the array of the average contributions over all index values $i$ of the time-series array for which $\Delta n$ predictions can be obtained:
\begin{equation} \label{eq:EQ009}
S (k) = \frac{1}{N_c} \sum_{i = 1 + (m-1) \nu}^{N - \Delta n} \ln \left( Q_i (k) \right) \, \, \, ,
\end{equation}
where $N_c$ denotes the number of elements in the original time series for which $\mathscr{U}_i \neq \varnothing$. The quantities $S (k)$ are known as out-of-sample prediction-error arrays. (Although KS suggest the use of $N$ 
in Eq.~(\ref{eq:EQ009}), I believe that the use of $N_c$ is more convenient as it enables the direct comparison of the arrays obtained for different embedding dimensions.)
\end{itemize}
At this point, one remark is due. Technically, Eq.~(\ref{eq:EQ009}) is problematic from the strictly mathematical point of view: if, for some value of $k$, it so happens that $s_{ i + k } = \tilde{s}_{ j + k }$, 
$\forall j \in \mathscr{U}_i$, the logarithm is not defined. However, let me put aside such a possibility, anticipating that the `noise' present in the time-series measurements precludes such critical situations.

I would like to suggest one modification to the method put forward by KS: Eq.~(\ref{eq:EQ008}) could be replaced by the form:
\begin{equation} \label{eq:EQ009_1}
Q_i (k) = \left( \sum_{j \in \mathscr{U}_i} w_{ij} \right)^{-1} \sum_{j \in \mathscr{U}_i} w_{ij} \abs{ s_{ i + k } - \tilde{s}_{ j + k } } \, \, \, ,
\end{equation}
where the weights $w_{ij}$ take account of the proximity of the embedding vectors ${\ss}_i$ and $\tilde{\ss}_j$; the smaller the distance $d_{ij}$ between these two vectors, the larger the weight of their contribution to the 
prediction-error array $Q_i (k)$. In this context, the method described in Refs.~\cite{kantz1997} would correspond to $w_{ij} \equiv 1$, whereas the results of this work have been obtained using $w_{ij} = 1 - (d_{ij}/\epsilon)^2$; 
I have no reason to expect sizeable systematic effects, but I believe that the statistical weights $w_{ij}$ should be used. Obviously, the weight vanishes at $d_{ij} = \epsilon$ and, given the condition (\ref{eq:EQ007}), it 
may be thought of as vanishing for $d_{ij} > \epsilon$. Of course, one may use higher (even) powers in the definition of the weights (perhaps, KS avoid the introduction of weights in the determination of $Q_i (k)$ due to this 
arbitrariness); the power of $4$ was also used in this work, but the impact on the results was found insignificant.

In chaotic systems, $S (k)$ of Eq.~(\ref{eq:EQ009}) should be linearly increasing with $k$, up to the point where it saturates to the average absolute distance of two arbitrary embedding vectors on the attractor. KS emphasise 
the importance of the linearity of $S (k)$ (with $k$) at small $k$ values; a positive slope seen on the ($k$,$S (k)$) scatter plot is indicative of chaotic behaviour.

\begin{figure}
\begin{center}
\includegraphics [width=15.5cm] {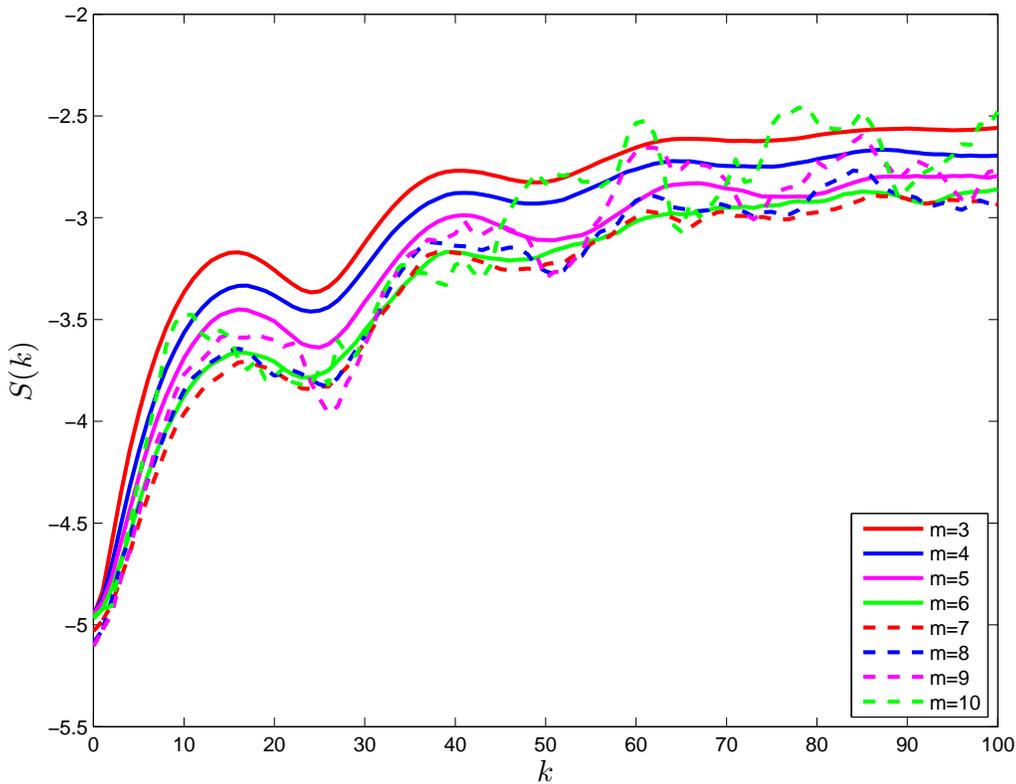}
\caption{\label{fig:SDn0p0170}The out-of-sample prediction-error arrays $S (k)$ (for the details, see Section \ref{sec:Lyapunov}) corresponding to the neighbourhood size $\epsilon = 1.70 \cdot 10^{-2}$.}
\vspace{0.35cm}
\end{center}
\end{figure}

Figure \ref{fig:SDn0p0170} shows all out-of-sample prediction-error arrays $S (k)$ corresponding to $\epsilon = 1.70 \cdot 10^{-2}$ (those not shown did not fulfil the acceptance criterion of the $10$ non-zero contributions 
to $S (k)$). The plot is representative of the general behaviour of $S (k)$. The undulation of $S (k)$ is associated with the periodicity of the input data series. A linear segment in the ($k$,$S (k)$) scatter plot is rather 
difficult to find. On the other hand, it is evident that $S (k)$ increases with $k$ (below about $40-50$ sampling intervals) and saturates around $70$ sampling intervals into the future. The outcome of the analysis of the 
luminosity measurements of PG 1159-035 resembles the one obtained by KS from far-infrared laser data (data sets SF\_A and SF\_Acont in Ref.~\cite{timeseries}).

One way to cope with the undulations of $S (k)$ would be to analyse the local extrema and derive limits for the maximal Lyapunov exponent; however, such an analysis would rest upon very few data points. A better option is to 
fit a suitable monotonic function to the $S (k)$ values and capture the general trend of $S (k)$; this option will be followed in the remaining part of this section.

For the purpose of fitting to the $S (k)$ arrays, the MINUIT package \cite{jms} of the CERN library (FORTRAN version) was used. Each optimisation was achieved with the (robust) sequence: SIMPLEX, MINIMIZE, MIGRAD, and MINOS.
\begin{itemize}
\item SIMPLEX is a function-minimisation method, using the simplex method of Nelder and Mead. Being a stepping method, SIMPLEX does not produce a Hessian matrix.
\item MINIMIZE minimises the user-defined function by calling MIGRAD and reverts to SIMPLEX in case that the MIGRAD call fails to converge.
\item MIGRAD is the workhorse of the MINUIT software library. It is a variable-metric method, also checking for the positive-definiteness of the Hessian matrix.
\item MINOS performs a detailed error analysis for each of the model parameters separately. It may be time consuming, but its results (i.e., the asymmetric uncertainties of the model parameters) are reliable as they take the 
non-linearities into account, as well as the correlations among the model parameters.
\end{itemize}
All the aforementioned methods admit an optional argument, limiting the maximal number of calls of the method; if this number is reached, that method is terminated (by MINUIT, internally) regardless of whether it has converged 
or not. To ensure the successful termination of the MINUIT application and the convergence of its methods, the output of the application was automatically displayed and checked for failures.

The original out-of-sample prediction-error arrays $S (k)$, obtained by the appropriate variation of the quantities $m$ and $\epsilon$ (i.e., within the linearity region in the ($\ln \epsilon$,$\ln C(\epsilon)$) scatter plots), 
were fitted to by the function
\begin{equation} \label{eq:EQ010}
S (k) = \ln \left[ x_1 \exp\left[ x_2 \left( 1 + \frac{x_3}{x_1} \right) k \right] + x_3 \right] \, \, \, ,
\end{equation}
where the parameters $x_{1,2,3}$ are associated with the variation of $S (k)$ between $k=0$ and saturation, the maximal Lyapunov exponent ($\lambda$), and the saturation level of $S (k)$, respectively; the expansion of $S (k)$ 
of Eq.~(\ref{eq:EQ010}) for small $k$ values is: $S (k) \approx \ln (x_1 + x_3) + x_2 k$. Other suitable three-parameter forms were also tried, but generally gave inferior $\chi^2$ results. Nevertheless, results were also 
obtained (for the sake of comparison) with one of these alternative forms, namely
\begin{equation} \label{eq:EQ011}
S (k) = x_3 - \frac{x_2 x_1^2}{k - x_1} \, \, \, ,
\end{equation}
where the meaning of the parameters $x_2$ and $x_3$ is the same as for the form (\ref{eq:EQ010}), whereas $x_1<0$ in the case of the form (\ref{eq:EQ011}) represents the position of the vertical asymptote of the hyperbolic form; 
the expansion of $S (k)$ of Eq.~(\ref{eq:EQ011}) for small $k$ values is: $S (k) \approx x_3 + x_1 x_2 + x_2 k$.

\begin{figure}
\begin{center}
\includegraphics [width=15.5cm] {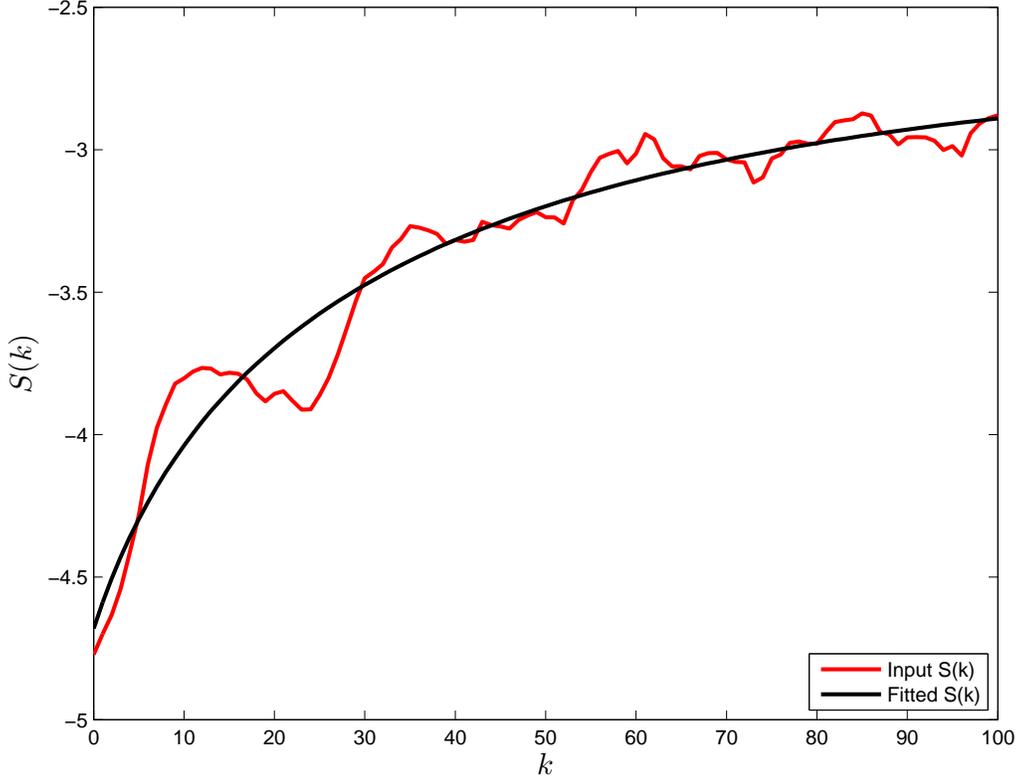}
\caption{\label{fig:SDnFitted}Original and fitted values of an out-of-sample prediction-error array $S (k)$, see Eq.~(\ref{eq:EQ009}). The figure corresponds to $m=10$ and $\epsilon = 2.10 \cdot 10^{-2}$, which yields a 
$\chi^2$ value close to the median value obtained in the original fits to the $S (k)$ arrays.}
\vspace{0.35cm}
\end{center}
\end{figure}

A constant working uncertainty of $0.1$ was assigned to each input value. Given the redefinition of the fitted uncertainties, taking account of the quality of each fit via the application of the Birge factor $\sqrt{\chi^2/{\rm NDF}}$ 
(NDF stands for the number of degrees of freedom in the fit, namely the number of input data points reduced by the number of the fit parameters, i.e., $3$), the value of the assigned uncertainty is irrelevant, i.e., the choice 
of another (non-zero) value leads to identical results. The optimisation application did not terminate successfully in $4$ (out of $140$) cases, which (of course) had to be excluded. (No failures were found when the data were 
fitted to by the form of Eq.~(\ref{eq:EQ011}).) The results of a representative fit (original and fitted $S (k)$ data) are shown in Fig.~\ref{fig:SDnFitted}.

To obtain reliable estimates for $\lambda$, the results of the fits were processed further. The minimal value of $\chi^2$, obtained from these fits, was equal to $46.02$ (for $97$ degrees of freedom), whereas the maximal one 
was $1539.73$. Inspection of the $\chi^2$ histogram demonstrated that, though reasonable values were obtained in most cases, unreasonably large $\chi^2$ results were generally obtained when the acceptance criterion of the $10$ 
non-zero contributions to $S (k)$ was barely fulfilled. The average $\chi^2$ was equal to about $187.53$, whereas the median value was $120.90$. To be rid of the cases with unreasonably large $\chi^2$ values, an acceptance 
criterion was introduced, at twice the median $\chi^2$ value of the original distribution. As a result, fits were accepted only if $\chi^2 \lesssim 241.80$. Discarded were $22$ (out of the original $136$) fits; the remaining 
cases were processed further, yielding the main results of this work.

\begin{figure}
\begin{center}
\includegraphics [width=15.5cm] {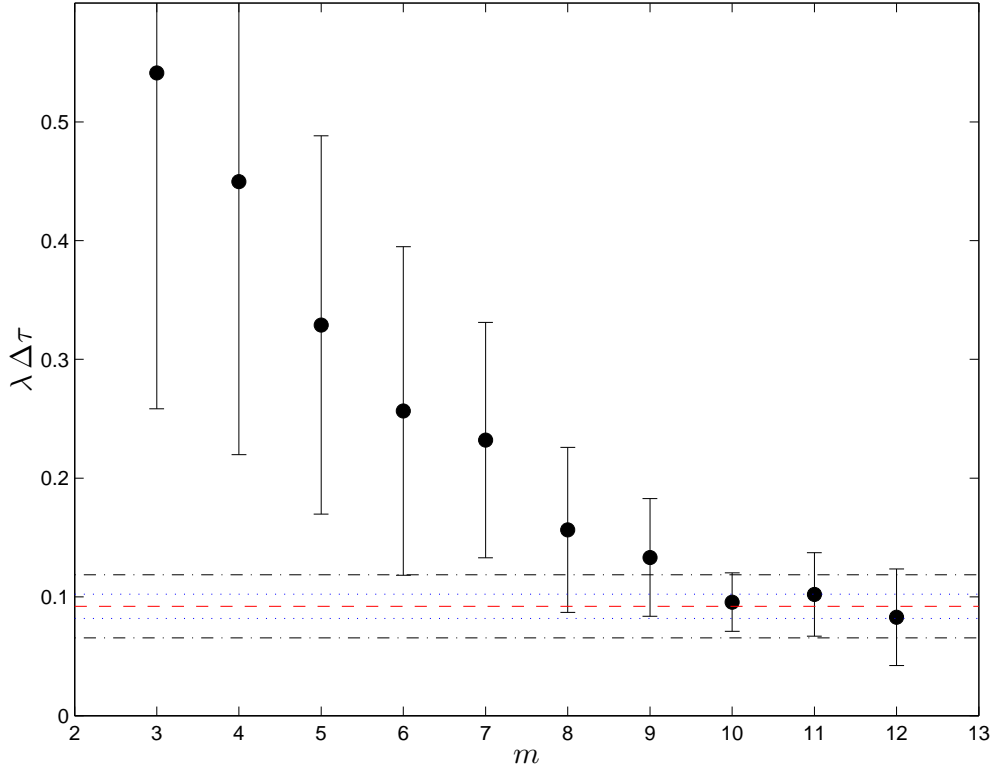}
\caption{\label{fig:Lyapunov}The maximal Lyapunov exponents extracted from the filtered luminosity measurements of PG 1159-035 using Eq.~(\ref{eq:EQ010}). The sum of the statistical and systematic uncertainties is shown for 
each data point; the statistical uncertainties have been corrected for the quality of each fit. The red dashed line represents the grand mean over the $\epsilon$ values for the sufficient embeddings ($10 \leq m \leq 12$). The 
blue dotted lines represent the $1 \sigma$ limits of the statistical uncertainty of the grand mean, whereas the black dash-dotted lines correspond to the ($1 \sigma$ limits of the) systematic uncertainty.}
\vspace{0.35cm}
\end{center}
\end{figure}

The maximal Lyapunov exponents extracted from the filtered luminosity measurements of PG 1159-035 for $3 \leq m \leq 12$ are shown in Fig.~\ref{fig:Lyapunov}. Noticeable in this figure is the decrease of the uncertainties with 
increasing embedding dimension; this is the result of the better compatibility of the extracted $\lambda$ values for the different neighbourhood sizes as the embedding dimension approaches the optimal one. The values should be 
compatible for all sufficient embeddings, and indeed they are. One conclusion may be drawn from this plot: the maximal Lyapunov exponent is positive. Restricting the analysis only to sufficient embeddings ($10 \leq m \leq 12$), 
one obtains $\lambda = (9.2 \pm 1.0 ({\rm stat.}) \pm 2.7 ({\rm syst.})) \cdot 10^{-2}~\Delta \tau^{-1}$. The first uncertainty is statistical (average over fitted uncertainties, corrected for the quality of each fit), 
the second systematic (reflecting the variation of $\lambda$ with $\epsilon$ for the sufficient embeddings). (The use of the hyperbolic form of Eq.~(\ref{eq:EQ011}) in the fits yielded a compatible result, namely 
$\lambda = (8.3 \pm 1.0 ({\rm stat.}) \pm 2.3 ({\rm syst.})) \cdot 10^{-2}~\Delta \tau^{-1}$.)

\subsection{\label{sec:L2Norm}Changes in the important results when the $L^2$-norm distance is used}

The results of Sections \ref{sec:CorrelDim} and \ref{sec:Lyapunov}, obtained with the use of the $L^\infty$ distance, are the main results of this work. Assessed in this section is the importance of the changes when using a 
different norm in the quantification of the closeness of two embedding vectors, namely the $L^2$ norm (Euclidean distance). At fixed $\epsilon$, the change from the $L^\infty$ to the $L^2$ distance leads to the extraction of 
a smaller number of neighbours, hence contributions to the correlation sums and to the out-of-sample prediction-error arrays $S (k)$. Therefore, the increase of the uncertainty of the extracted estimates for the maximal 
Lyapunov exponent is expected.

The linearity between $\ln C(\epsilon)$ and $\ln \epsilon$ was investigated as in Section \ref{sec:CorrelDim}. As in the case of the $L^\infty$ distance, the out-of-sample prediction-error arrays $S (k)$ were created and 
further analysed only if at least $10$ neighbours could be found in the database. Table \ref{tab:DimensionL2} is the equivalent of Table \ref{tab:DimensionLinf} in the case of the $L^2$-norm distance. One notices that the 
slope $\alpha$ does not saturate with increasing embedding dimension; at the present time, I cannot explain why.

\begin{table}
{\bf \caption{\label{tab:DimensionL2}}}The equivalent of Table \ref{tab:DimensionLinf} when the $L^2$-norm distance is used in the determination of the correlation sums.
\vspace{0.2cm}
\begin{center}
\begin{tabular}{|c|c|c|c|c|c|c|}
\hline
$m$ & $\epsilon_{\rm min}$ & $\epsilon_{\rm max}$ & $\alpha$ & $\delta \alpha$ & $\beta$ & $\delta \beta$\\
\hline
\hline
$3$ & $6.00 \cdot 10^{-3}$ & $2.10 \cdot 10^{-2}$ & $2.9683$ & $0.0035$ & $5.445$ & $0.014$\\
$4$ & $6.00 \cdot 10^{-3}$ & $2.50 \cdot 10^{-2}$ & $3.9191$ & $0.0071$ & $7.398$ & $0.028$\\
$5$ & $6.00 \cdot 10^{-3}$ & $2.70 \cdot 10^{-2}$ & $4.884$ & $0.016$ & $9.430$ & $0.059$\\
$6$ & $9.00 \cdot 10^{-3}$ & $3.00 \cdot 10^{-2}$ & $5.978$ & $0.029$ & $11.82$ & $0.11$\\
$7$ & $1.30 \cdot 10^{-2}$ & $3.20 \cdot 10^{-2}$ & $7.021$ & $0.056$ & $14.03$ & $0.20$\\
$8$ & $1.90 \cdot 10^{-2}$ & $3.70 \cdot 10^{-2}$ & $7.848$ & $0.081$ & $15.40$ & $0.27$\\
$9$ & $2.50 \cdot 10^{-2}$ & $3.90 \cdot 10^{-2}$ & $9.99$ & $0.21$ & $20.99$ & $0.68$\\
$10$ & $2.90 \cdot 10^{-2}$ & $4.30 \cdot 10^{-2}$ & $11.04$ & $0.25$ & $22.88$ & $0.81$\\
$11$ & $3.40 \cdot 10^{-2}$ & $4.50 \cdot 10^{-2}$ & $11.41$ & $0.33$ & $22.7$ & $1.0$\\
$12$ & $3.80 \cdot 10^{-2}$ & $4.50 \cdot 10^{-2}$ & $12.80$ & $0.37$ & $25.7$ & $1.2$\\
\hline
\end{tabular}
\end{center}
\vspace{0.5cm}
\end{table}

The out-of-sample prediction-error arrays $S (k)$ were obtained and fitted to, as described in Section \ref{sec:Lyapunov}. The maximal Lyapunov exponent was extracted using Eq.~(\ref{eq:EQ010}) for all embedding dimensions and 
$\epsilon_{\rm min} \leq \epsilon \leq \epsilon_{\rm max}$ (linearity region in the ($\ln \epsilon$,$\ln C(\epsilon)$) plane, see Table \ref{tab:DimensionL2}). An upper cut was applied to the resulting $\chi^2$ values as in 
Section \ref{sec:Lyapunov}. The final outcome is: $\lambda=(16.4 \pm 2.0 ({\rm stat.}) \pm 5.0 ({\rm syst.})) \cdot 10^{-2}~\Delta \tau^{-1}$, larger than, yet not incompatible with, the more accurate result obtained with the 
$L^\infty$ distance.

\section{\label{sec:Conclusions}Discussion and conclusions}

The goal of this work was the analysis of the luminosity measurements of the pre-white dwarf PG 1159-035, in fact those of the measurements which found their way to the `Time-series data source Archives: Santa F\'e Time Series 
Competition' \cite{timeseries}. It is my belief that the set of the experimental data, acquired in the 1989 runs of the Whole Earth Telescope (WET) project, is considerably more extensive.

Following the results of Refs.~\cite{winget1991,costa2008}, the seventeen available time series were suitably band-passed using an elliptical filter, whose $13$ recursion coefficients are listed in Table \ref{tab:RecursionCoefficients}. 
The filtered data was split into two parts of comparable sizes, one yielding the training (learning) set or database, the second the test set. The optimal embedding dimension was determined using Cao's method \cite{cao1997}, see 
Section \ref{sec:Cao}: it appears that optimal embeddings require a $10$-dimensional space. This choice was confirmed in Section \ref{sec:CorrelDim} by an analysis of the correlation dimension.

The extraction of the maximal Lyapunov exponent $\lambda$ was next attempted by fitting a monotonic function (see Eq.~(\ref{eq:EQ010})) to the out-of-sample prediction-error arrays $S (k)$, defined in Eq.~(\ref{eq:EQ009}); 
the original arrays contain sizeable undulations, hindering the determination of a region in the ($k$,$S (k)$) plane within which a linear relationship (i.e., the signature of a chaotic dynamical system) holds. A modification 
is proposed in this work in the evaluation of the $S (k)$ arrays, taking account of the distance between the specific embedding vector of the test set and its corresponding partner in the training set: the smaller the distance 
between these two vectors, the larger the weight assigned to the associated predictions in the determination of the out-of-sample prediction-error arrays $S (k)$, see Eq.~(\ref{eq:EQ009_1}).

The data analysis suggests that the maximal Lyapunov exponent $\lambda$, associated with the luminosity measurements of PG 1159-035, is equal to $(9.2 \pm 1.0 ({\rm stat.}) \pm 2.7 ({\rm syst.})) \cdot 10^{-2}~\Delta \tau^{-1}$, 
where $\Delta \tau$ represents the sampling interval in the measurements ($10$ s). It was found that the extracted $\lambda$ values do not show significant dependence on the embedding dimension for sufficient embeddings 
($10 \leq m \leq 12$), see Fig.~\ref{fig:Lyapunov}. The findings of this work suggest that it is very likely that the source of the observations is indeed chaotic.

The aforementioned results refer to the use of the $L^\infty$-norm distance. Interestingly, the use of the Euclidean ($L^2$-norm) distance yields a larger, yet not incompatible, result for the maximal Lyapunov exponent.

This work is the first step I have taken in analysing the luminosity measurements of the pre-white dwarf PG 1159-035. I would be glad to receive the entirety of the data (1989 runs) from a credible source, e.g., directly from 
one of the members of the WET Collaboration. The analysis of the data after suitably involving all files in both the training and the test sets is currently under investigation \cite{matsinos2018}.

\begin{ack}
I acknowledge an e-mail exchange with S.~Hollos.

All the figures have been created with MATLAB \textregistered~(The MathWorks, Inc., Natick, Massachusetts, United States).
\end{ack}

\end{document}